\documentclass[12pt]{article}
\catcode`\@=11
\@addtoreset{equation}{section}

\global\arraycolsep=2pt 
\newcommand{\cO}{{\mathcal O}}

\newcommand{\ga}{\gamma}

\newcommand{\la}{\lambda}

\newcommand{\str}{{\rm STr}}

\newcommand{\alg}[1]{\mathfrak{#1}}

\newcommand{\nln}{\nonumber\\}

\newcommand{\no}{\nonumber}
\newcommand{\mL}{\mathcal L}
\newcommand{\oinv}{\mathcal{O}^{{\rm inv}}}
\newcommand{\gP}{{\rm\bf P}}
\newcommand{\gJ}{{\rm\bf J}}
\newcommand{\gQ}{{\rm\bf Q}}
\newcommand{\cM}{{\mathcal M}}

\newcommand{\bth}{\bar{\theta}}
\newcommand{\dzero}{_{(0)}}

\newcommand{\dtwo}{_{(2)}}

\usepackage{mathrsfs,amsbsy,amssymb,latexsym,amsfonts,amsmath,cite}
\usepackage{graphicx,color}
\usepackage{mathtools}

\allowdisplaybreaks

\begin{document}

\begin{flushright}
\parbox{4cm}
{KUNS-2624 \\ 
\today }
\end{flushright}

\vspace*{1.5cm}

\begin{center}
{\Large \bf 
Supercoset construction of Yang-Baxter deformed AdS$_5\times$S$^5$ backgrounds}
\vspace*{1.5cm}\\
{\large Hideki Kyono\footnote{E-mail:~h\_kyono@gauge.scphys.kyoto-u.ac.jp}
and Kentaroh Yoshida\footnote{E-mail:~kyoshida@gauge.scphys.kyoto-u.ac.jp}} 
\end{center}

\vspace*{0.5cm}

\begin{center}
{\it Department of Physics, Kyoto University, \\ 
Kitashirakawa Oiwake-cho, Kyoto 606-8502, Japan} 
\end{center}

\vspace{1cm}

\begin{abstract}
We proceed to study Yang-Baxter deformations of the AdS$_5\times$S$^5$
superstring with the classical Yang-Baxter equation. 
We make a general argument on the supercoset construction 
and present a formula to describe the dilaton in terms of classical $r$-matrices.
The supercoset construction is explicitly performed for some classical $r$-matrices   
and the full backgrounds including the Ramond-Ramond (R-R) sector and dilaton are derived. 
Within the class of abelian $r$-matrices, the perfect agreement is shown for well-known examples 
including gravity duals of non-commutative gauge theories, $\gamma$-deformations of S$^5$ 
and Schr\"odinger spacetimes. 
It is remarkable that the supercoset construction works well, 
even if the resulting backgrounds are not maximally supersymmetric. 
In particular, three-parameter $\gamma$-deformations of S$^5$ and 
Schr\"odinger spacetimes do not preserve any supersymmetries. 
As for non-abelian $r$-matrices, we will focus upon  
a specific example. The resulting background does not satisfy the equation of motion 
of the Neveu-Schwarz-Neveu-Schwarz (NS-NS) two-form because the R-R three-form is not closed. 

\end{abstract}

\setcounter{footnote}{0}
\setcounter{page}{0}
\thispagestyle{empty}

\newpage

\tableofcontents

\section{Introduction}

The Yang-Baxter deformation \cite{Klimcik} is a systematic way to study integrable deformations 
of non-linear sigma models in two dimensions. Given a classical $r$-matrix satisfying the classical 
Yang-Baxter equation (CYBE), an integrable deformation is determined and the associated Lax pair 
follows automatically. This correspondence between a deformed geometry and 
a classical $r$-matrixindicates a profound connection between a differential geometry and 
a finite-size matrix. Hence it is significant to make the understanding of the Yang-Baxter deformation 
much deeper from the viewpoints of theoretical physics and pure mathematics. 

\medskip 

The Yang-Baxter deformation was originally invented for principal chiral models 
with the modified classical Yang-Baxter equation (mCYBE). 
Now that it is generalized to symmetric cosets \cite{DMV} and the homogeneous CYBE \cite{MY-YBE},  
one can study Yang-Baxter deformations of symmetric coset sigma models 
with a lot of examples of classical $r$-matrices. For the related affine algebras, 
see the series of works \cite{KMY-alg,KOY,KY-Sch}.

\medskip

The most interesting coset sigma model is type IIB string theory on AdS$_5\times$S$^5$ 
in the context of the anti-de Sitter/conformal field theory (AdS/CFT) correspondence \cite{M}. 
The classical string action has been constructed in the Green-Schwarz formulation 
based on a supercoset \cite{MT}
\[
\frac{PSU(2,2|4)}{SO(1,4) \times SO(5)}\,. 
\] 
This coset enjoys the $\mathbb{Z}_4$-grading property 
and ensures classical integrability \cite{BPR} 
(for a nice review, see \cite{AF-review}). 
The integrability plays an important role in checking 
the conjectured relation in AdS/CFT   
(for a comprehensive review, see \cite{review}).

\medskip 

By employing the Yang-Baxter deformation, Delduc, Magro and Vicedo constructed 
the classical action of a $q$-deformed AdS$_5\times$S$^5$ superstring \cite{DMV2}. 
This deformation comes from the classical $r$-matrix of Drinfel'd-Jimbo type 
satisfying the mCYBE \cite{DJ}. The string-frame metric and Neveu-Schwarz-Neveu-Schwarz 
(NS-NS) two-form were derived 
by Arutyunov, Borsato and Frolov \cite{ABF}. Then they performed the supercoset construction 
and derived the remaining sector \cite{ABF2} (for earlier attempts, see \cite{HRT,LRT}). 
As a result, the full background does not satisfy the equations of motion of type IIB supergravity, 
although it is related to a complete solution \cite{HT-sol} via T-dualities 
apart from the dilaton part. In particular, the dilaton cannot be separated so that 
the Ramond-Ramond (R-R) flux should satisfy the Bianchi identity. 
Recently, Arutyunov et al.\ proposed an exciting conjecture that type IIB supergravity itself 
would get deformed, for example, the definition of R-R field strength may be modified 
\cite{scale}. This ``modified gravity conjecture'' may be connected to our result presented here. 

\medskip 

One may also consider Yang-Baxter deformations of the AdS$_5\times$S$^5$ superstring 
with classical $r$-matrices satisfying the homogeneous CYBE \cite{KMY-Jordanian-typeIIB}. 
A strong advantage in this case is that partial deformations of AdS$_5\times$S$^5$ are possible. 
In fact, for well-known backgrounds including gravity duals of noncommutative gauge theories \cite{HI,MR}, 
$\gamma$-deformations of S$^5$ \cite{LM,Frolov}, Schr\"odinger spacetimes \cite{MMT}, 
the associated classical $r$-matrices have been identified in a series of works 
\cite{LM-MY,MR-MY,Sch-MY,KMY-SUGRA,MY-duality,Stijn1,Stijn2} 
(for short summaries, see \cite{MY-summary}). 
However, the analysis has been limited to the bosonic sector so far, and it is still necessary 
to confirm the R-R sector and dilaton by performing the supercoset construction explicitly. 

\medskip 

The goal of this present work is to perform the supercoset construction 
and present the resulting backgrounds for some classical $r$-matrices. 
We will first give a general treatment basically by following the seminal 
paper by Arutyunov, Borsato and Frolov \cite{ABF2}. Then we derive the backgrounds 
for some classical $r$-matrices. As a byproduct, we present the master formula to 
describe the dilaton in terms of classical $r$-matrices. 

\medskip

Within the class of abelian classical $r$-matrices, 
the perfect agreement is shown for well-known examples including 
gravity duals of non-commutative gauge theories \cite{HI,MR}, 
$\gamma$-deformations of S$^5$ \cite{LM,Frolov},
and Schr\"odinger spacetimes \cite{MMT}. 
It is worth noting that the supercoset construction works well, 
even though the resulting backgrounds are not maximally supersymmetric. 
More strikingly, three-parameter $\gamma$-deformations of S$^5$ 
and Schr\"odinger spacetimes do not preserve any supersymmetries \cite{MMT}. 
Hence it seems likely that the supercoset construction 
works well with the class of the abelian classical $r$-matrices.  
It is consistent with the interpretation as TsT transformations 
\cite{Frolov,AAF,KY-Sch,Benoit,Stijn2,KKSY}. 

\medskip

As for non-abelian classical $r$-matrices, we will focus on 
a specific example discussed in \cite{KMY-SUGRA,MY-duality}. 
The resulting background does not satisfy the equation of motion 
of the NS-NS two-form because the Bianchi identity of the R-R three-form is broken, 
namely the field strength is not closed. 
It is also remarkable that this background is different from 
the one proposed in \cite{KMY-SUGRA,MY-duality} 
and hence the identification made in \cite{KMY-SUGRA,MY-duality} was not correct. 
Anyway, this result indicates that 
there would be some potential problems in the non-abelian cases. 
This is really intriguing but just an example. It is of importance  
to study extensively other non-abelian $r$-matrices. 

\medskip 

This paper is organized as follows. 
Section 2 introduces the classical action of Yang-Baxter deformed AdS$_5\times$S$^5$ 
superstring based on the CYBE. In Sect.\ 3, we discuss the supercoset construction, 
by following the procedure of \cite{ABF2}. Most of the argument does not 
rely on specific expressions of classical $r$-matrices and is quite general. 
We present the conjectured master formula to describe the dilaton in terms of classical $r$-matrices. 
In Sect.\ 4, we present the resulting backgrounds for concrete examples of classical $r$-matrices. 
Section 5 is devoted to the conclusion and discussion. 
Appendix A provides a matrix representation of the superalgebra $\mathfrak{su}(2,2|4)$.

\section{Yang-Baxter deformed AdS$_5\times$S$^5$ superstring}

In this section, we give a short introduction to the classical action of 
Yang-Baxter deformed AdS$_5\times$S$^5$ superstring based 
on the homogeneous CYBE \cite{KMY-Jordanian-typeIIB}. 
This construction basically follows from the work with the mCYBE \cite{DMV2}.   

\medskip

The deformed classical action of the AdS$_5\times$S$^5$ superstring is given by
\begin{eqnarray}
S=-\frac{\sqrt{\lambda_{\rm c}}}{4}\int_{-\infty}^\infty d\tau\int_0^{2\pi}d\sigma\,
(\ga^{ab}-\epsilon^{ab})\,
{\rm STr}\Bigl[A_a\, d\circ\frac{1}{1-\eta R_g\circ d}(A_b)\Bigr]\,, 
\label{YBsM}
\end{eqnarray}
where the left-invariant one-form $A_a$ is defined as
\begin{eqnarray}
A_a \equiv -g^{-1}\partial_a g\,,\quad\quad g\in SU(2,2|4) 
\end{eqnarray}
with the world-sheet index $a=(\tau,\sigma)$. 
Here, the conformal gauge is supposed and the world-sheet 
metric is taken to be the diagonal form $\ga^{ab}={\rm diag}(-1,+1)$. 
Hence there is no coupling of the dilaton to 
the world-sheet scalar curvature. 
The anti-symmetric tensor $\epsilon^{ab}$ is normalized as $\epsilon^{\tau\sigma}=+1$. 
The constant $\lambda_{\rm c}$ in front of the action (\ref{YBsM}) is the 't Hooft coupling.  
The deformation is measured by a constant parameter $\eta$, 
and the undeformed AdS$_5\times$S$^5$ action \cite{MT} is reproduced when $\eta=0$.

\medskip

A key ingredient in Yang-Baxter deformations is the operator $R_g$ defined as
\begin{eqnarray}
R_g(X)\equiv g^{-1}R(gXg^{-1})g\,, \quad\quad X\in \alg{su}(2,2|4)\,,
\label{R}
\end{eqnarray}
where a linear operator $R:\alg{su}(2,2|4)\to \alg{su}(2,2|4)$ 
is a solution of the CYBE\footnote{
In the original work \cite{KMY-Jordanian-typeIIB}, a wider class of $R$-operators is argued and 
their image is given by $\mathfrak{gl}(4|4)$.  
The $\mathfrak{gl}(4|4)$ image is restricted on $\alg{su}(2,2|4)$ 
under the coset projection $d$, as pointed out in \cite{Stijn1}.},
\begin{eqnarray}
[R(X),R(Y)]-R([R(X),Y]+[X,R(Y)])=0\,. \label{CYBE}
\end{eqnarray}
This $R$-operator is connected to a {\it skew-symmetric} classical $r$-matrix 
in the tensorial notation through the following formula: 
\begin{eqnarray}
R(X)=\text{STr}_2[r(1\otimes X)]=\sum_i\left( a_i\,\text{STr}[b_iX]-b_i\,\text{STr}[a_iX]\right)\,.
\label{R-r}
\end{eqnarray}
Here, $r$ is represented by 
\begin{eqnarray}
r=\sum_ia_i \wedge  b_i\equiv\sum_i \left(a_i \otimes b_i - b_i \otimes a_i\right)\qquad 
\mbox{with}\qquad a_i,~b_i\in\mathfrak{su}(2,2|4)\,.
\end{eqnarray}

\medskip

The projection operator $d$ is defined as 
\begin{eqnarray}
d &\equiv& P_1+2P_2-P_3\,, 
\label{op-d}
\end{eqnarray}
where $P_{\ell}\,(\ell=0,1,2,3)$ are projections to the $\mathbb{Z}_4$-graded components 
of $\mathfrak{su}(2,2|4)$. In particular, $P_0(\mathfrak{su}(2,2|4))$ is 
a local symmetry of the classical action, $\mathfrak{so}(1,4)\oplus\mathfrak{so}(5)$.
The numerical coefficients in the linear combination (\ref{op-d}) are fixed 
by requiring kappa symmetry \cite{MT,KMY-Jordanian-typeIIB}.

\section{Supercoset construction}

In this section, we shall consider the supercoset construction, 
starting from the deformed action (\ref{YBsM}). The following argument will be undertaken
without fixing a specific expression of classical $r$-matrices and hence will be quite general. 
Our purpose here is to extract the R-R fluxes and dilaton, and hence we will investigate 
the deformed action at the quadratic level of fermions.

\subsection{The $\mathfrak{su}(2,2|4)$ superalgebra}

For the subsequent argument, it is necessary to determine our convention and notation for
the $\mathfrak{su}(2,2|4)$ superalgebra. 
Hereafter, we will work with the following algebra \cite{MT}:   
\begin{eqnarray}
[\gP_{\check{m}},\gP_{\check{n}}] &=& \gJ_{\check{m}\check{n}}\,,
\quad [\gP_{\hat{m}},\gP_{\hat{n}}] = -\gJ_{\hat{m}\hat{n}}\,,\nonumber \\
\left[\gP_{\check{m}},\gJ_{\check{n}\check{p}}\right] 
&=&\eta_{\check{m}\check{n}}\,\gP_{\check{p}}-\eta_{\check{m}\check{p}}\,\gP_{\check{n}}\,,
\quad \left[\gP_{\hat{m}},\gJ_{\hat{n}\hat{p}}\right] 
= \eta_{\hat{m}\hat{n}}\,\gP_{\hat{p}}-\eta_{\hat{m}\hat{p}}\,\gP_{\hat{n}}\,, \nonumber \\
\left[\gJ_{\check{m}\check{n}},\gJ_{\check{p}\check{q}}\right] 
&=& \eta_{\check{n}\check{p}}\,\gJ_{\check{m}\check{q}}+(\text{3 terms})\,,
\quad \left[\gJ_{\hat{m}\hat{n}},\gJ_{\hat{p}\hat{q}}\right]
= \eta_{\hat{n}\hat{p}}\,\gJ_{\hat{m}\hat{q}}+(\text{3 terms})\,, \nonumber \\ 
\left[\gQ^I,\gP_{\check{m}} \right] 
&=& -\frac{i}{2}\epsilon^{IJ}\,\gQ^J \gamma_{\check{m}}\,, 
\quad \left[\gQ^I,\gP_{\hat{m}} \right] 
= \frac{1}{2}\epsilon^{IJ}\,\gQ^J \gamma_{\hat{m}}\,, \nonumber \\
\left[\gQ^I,\gJ_{\check{m}\check{n}} \right] 
&=& -\frac{1}{2}\delta^{IJ}\,\gQ^J \gamma_{\check{m}\check{n}}\,, 
\quad \left[\gQ^I,\gJ_{\hat{m}\hat{n}} \right] 
= -\frac{1}{2}\delta^{IJ}\,\gQ^J \gamma_{\hat{m}\hat{n}}\,, \nonumber \\
\bigl\{(\gQ^{\check{\alpha}\hat{\alpha}})^I,(\gQ^{\check{\beta}\hat{\beta}})^J\bigr\} 
&=& \delta^{IJ}\Bigl[ i K^{\check{\alpha}\check{\gamma}}
K^{\hat{\alpha}\hat{\beta}}{(\gamma^{\check{m}})_{\check{\gamma}}}^{\check{\beta}}\,
\gP_{\check{m}} - K^{\check{\alpha}\check{\beta}}K^{\hat{\alpha}\hat{\gamma}}
(\gamma^{\hat{m}})_{\hat{\gamma}}^{~\hat{\beta}}\,\gP_{\hat{m}} 
- \frac{i}{2}K^{\check{\alpha}\check{\beta}}K^{\hat{\alpha}\hat{\beta}}\,
{\bf 1}_{\bf 8}\Bigr] \no \\ 
&& -\frac{1}{2}\epsilon^{IJ}
\Bigl[ K^{\check{\alpha}\check{\gamma}}
K^{\hat{\alpha}\hat{\beta}}{(\gamma^{\check{m}\check{n}})_{\check{\gamma}}}^{\check{\beta}}\,
\gJ_{\check{m}\check{n}} - K^{\check{\alpha}\check{\beta}}K^{\hat{\alpha}\hat{\gamma}}
(\gamma^{\hat{m}\hat{n}})_{\hat{\gamma}}^{~\hat{\beta}}\,\gJ_{\hat{m}\hat{n}}\Bigr]
\,. \label{algebra}
\end{eqnarray}
The generators $\gP_m$ are translations and 
the index $m=(\check{m},\hat{m})~(\check{m}=0,\ldots,4;~\hat{m}=5,\ldots,9)$ 
describes the ten-dimensional spacetime, where the indices $\check{m}$ and $\hat{m}$ 
are for AdS$_5$ and S$^5$, respectively. Then $\gJ_{\check{m}\check{n}}$ and 
$\gJ_{\hat{m}\hat{n}}$ describe rotations in AdS$_5$ and S$^5$, respectively. 
The supercharges $\gQ^I~(I=1,2)$ are written as $\gQ^I=(\gQ^{\check{\alpha}\hat{\alpha}})^I~
(\check{\alpha}=1,\ldots,4; ~\hat{\alpha}=1,\ldots,4)$. 
The antisymmetric tensor $\epsilon^{IJ}~(I,J=1,2)$ is normalized as $\epsilon^{12}=+1$. 
The constant matrices $K^{\check{\alpha}\check{\beta}}$ and $K^{\hat{\alpha}\hat{\beta}}$   
are charge conjugation matrices in AdS$_5$ and S$^5$, respectively.

\subsection{A group parametrization and the left-invariant current}

Then let us introduce a parametrization of the group element 
$g \in SU(2,2|4)$ as follows: 
\begin{eqnarray}
g=g_{\rm b}\,g_{\rm f}\,. \label{para1}
\end{eqnarray}
Herred{,} $g_{\rm b}$ is a bosonic element and parametrized with an appropriate coordinate system, 
depending on the backgrounds we are concerned with, as in the previous works \cite{MR-MY,LM-MY,Sch-MY}. 
We assume that the bosonic element is parametrized as 
\begin{eqnarray}
g_{\rm b} & = &g_{\rm b}{}^{\text{AdS}_5}\,g_{\rm b}{}^{\text{S}^5}\,,\no\\
g_{\rm b}{}^{\text{AdS}_5} &=&\exp \Bigl[\,x^0\,P_0+x^1\,P_1+x^2\,P_2+x^3\,P_3\,\Bigr]\,
\exp \Bigl[\,(\log\,z) \,D\,\Bigr]\,,\no\\
g_{\rm b}{}^{\text{S}^5}&=&
\exp \Bigl[\,\frac{i}{2}( \phi_1\, h_1+\phi_2\,h_2+\phi_3\,h_3)\,\Bigr]\,\exp \Bigl[\,\zeta\,\gJ_{68}\Bigr]
\,\exp\Bigl[-i\,r\,\gP_6\Bigr]\,.
\end{eqnarray}
Note here that the translations $P_\mu$\,, the dilatation $D$ and 
the Cartan generators of $\mathfrak{su}(4)$ $h_i~(i=1,2,3)$ are embedded into $8\times8$
matrices as
\begin{eqnarray}
\begin{pmatrix}
~P_\mu & ~\boldsymbol{0_4} \\  
~\boldsymbol{0_4} & ~\boldsymbol{0_4}
\end{pmatrix}\,,\quad
\begin{pmatrix}
~D & ~\boldsymbol{0_4} \\  
~\boldsymbol{0_4} & ~\boldsymbol{0_4}
\end{pmatrix}\,,\quad
\begin{pmatrix}
~\boldsymbol{0_4} & ~\boldsymbol{0_4} \\  
~\boldsymbol{0_4} & ~h_{i}
\end{pmatrix}\,. 
\end{eqnarray}
The coordinates $x^\mu$ and $z$ describe the Poincar\'e AdS$_5$,  
and $r,\,\zeta,\,\phi_{i=1,2,3}$ parametrize the round S$_5$. 
Then $g_{\rm f}$ is a group element generated by the supercharges as follows:  
\begin{eqnarray}
g_{\rm f} &=& \exp (\gQ^I {\theta_I}) \qquad (I=1,2)\,,  \nonumber \\ 
&& \mbox{where} \qquad \gQ^I \theta_I \equiv (\gQ^{\check{\alpha}\hat{\alpha}})^I\,
(\theta_{\check{\alpha}\hat{\alpha}})_I \qquad 
(\check{\alpha}=1,\ldots,4;~\hat{\alpha}=1,\ldots,4)\,. 
\end{eqnarray}
Here, $\theta_I=(\theta_{\check{\alpha}\hat{\alpha}})_I$ are Grassmann-odd coordinates and 
correspond to a couple of 16-component Majorana-Weyl spinors satisfying 
the Majorana condition:
\begin{eqnarray}
\bar{\theta}_I\equiv \theta^\dag_I \gamma^0={}^t\theta_I (K\otimes K)\,.
\end{eqnarray}

\medskip 

Then the left-invariant one-form $A$ can be expanded as \cite{MT}
\begin{eqnarray}
\label{currentA}
A &=& (e^m+\frac{i}{2}\bar{\theta}_I\gamma^m D^{IJ}\theta_J)\, {\gP}_m 
- \gQ^I\, D^{IJ}\theta_J+\frac{1}{2}\omega^{mn}\,\gJ_{mn}\no\\
&& \qquad - \frac{1}{4}\epsilon^{IJ}\bar{\theta}_I(\gamma^{\check{m}\check{n}}\,
\gJ_{\check{m}\check{n}} 
- \gamma^{\hat{m}\hat{n}}\,\gJ_{\hat{m}\hat{n}})D^{JK}\theta_K\,, 
\end{eqnarray}
where the covariant derivative for $\theta$ is defined as 
\begin{eqnarray}
D^{IJ}\theta_J = \delta^{IJ}\left( d\theta_J - \frac{1}{4}\omega^{mn}\gamma_{mn}\theta_J \right) 
+ \frac{i}{2}\epsilon^{IJ}\,e^m\,\gamma_m\theta_J\,. 
\end{eqnarray}
Here, the last term represents the contribution of the R-R five-form field strength. 

\medskip 

For later convenience, it is helpful to rearrange the above expansion of $A$ 
with respect to the order of $\theta$ as follows:  
\begin{eqnarray}
A&=&A_{(0)}+A_{(1)}+A_{(2)}\,. \no
\end{eqnarray}
Here, $A_{(p)}$ is the $p$-th order of $\theta$ and 
the explicit expressions of $A_{(p)}$ are given by 
\begin{eqnarray}
A_{(0)}&=&e^m\,{\gP}_m+\frac{1}{2}\omega^{mn}\,\gJ_{mn}\,,\no\\
A_{(1)}&=&-\gQ^I\,D^{IJ}\,\theta_J\,,\no\\
A_{(2)}&=&\frac{i}{2}\bar{\theta}_I\gamma^m D^{IJ}\theta_J\,{\gP}_m 
- \frac{1}{4}\epsilon^{IJ}\bar{\theta}_I(\gamma^{\check{m}\check{n}}\,\gJ_{\check{m}\check{n}}
-\gamma^{\hat{m}\hat{n}}\,\gJ_{\hat{m}\hat{n}})D^{JK}\theta_K\,. 
\end{eqnarray}
Thus we have prepared to write down the undeformed action of the AdS$_5\times$S$^5$ superstring 
at the quadratic order of $\theta$.

\subsection{Decomposing the deformation operator}

In the case of the deformed action (\ref{YBsM}), it is further necessary to expand  
the deformation operator in terms of $\theta$, because $R_g$ contains the adjoint operation 
with $g$ as denoted in (\ref{R}). We will basically follow the strategy of \cite{ABF2} hereafter.

\medskip

Let us introduce the following operator $\mathcal{O}$ and expand it in terms of $\theta$ as
\begin{eqnarray}
\cO&\equiv&1-\eta R_g\circ d\no\\
&=&\cO_{(0)}+\cO_{(1)}+\cO_{(2)}+\cO (\theta^3)\,. 
\end{eqnarray}
Then the inverse operator $\cO^{\rm inv}$ can also be expanded as 
\begin{eqnarray}
\cO^{{\rm inv}}&\equiv&\frac{1}{1-\eta R_g\circ d}\no\\
&=&\cO^{{\rm inv}}_{(0)}+\cO^{{\rm inv}}_{(1)}+\cO^{{\rm inv}}_{(2)}+\cO (\theta^3)\,. 
\label{perturbationO}
\end{eqnarray}
Here, due to the relation $\cO\circ \oinv=1$, each of the components $\cO^{\rm inv}_{(p)}~(p=0,1,2)$ 
can be expressed as follows: 
\begin{eqnarray}
\label{perturbationO2}
\cO^{{\rm inv}}_{(0)}&=&\frac{1}{1-\eta R_{g_b}\circ d}\,,\no\\
\oinv_{(1)}&=&-\oinv_{(0)}\circ \cO_{(1)}\circ\oinv_{(0)}\,,\no\\
\oinv_{(2)}&=&-\oinv_{(0)}\circ \cO_{(2)}\circ\oinv_{(0)}-\oinv_{(1)}\circ \cO_{(1)}\circ\oinv_{(0)}\,. 
\end{eqnarray}

In the following, for simplicity, we will concentrate only on
the bosonic deformations\footnote{It would also be 
interesting to consider fermionic deformations. 
For such an attempt, see \cite{KMY-Jordanian-typeIIB}.} 
generated by bosonic generators $a_i$, $b_i$ like 
\begin{eqnarray}
a_i\,,~b_i\,\in \mathfrak{su}(2,2)\oplus\mathfrak{su}(4)\,.
\end{eqnarray}
Then the action of $R_{g_{\rm b}}\circ d$ can be evaluated as 
\begin{eqnarray}
R_{g_{\rm b}}\circ d (\gP_m) 
= 2\left({\lambda_m}^n\gP_n+\frac{1}{2}{\lambda_m}^{np}\gJ_{np}\right)\,, 
\no \\
R_{g_{\rm b}}\circ d (\gJ_{mn})=0\,,\qquad R_{g_{\rm b}}\circ d (\gQ^I)=0\,.
\end{eqnarray}
Here, from the relation in (\ref{R-r}), $\lambda_m^{~~n}$ and $\lambda_m^{~~np}$ are 
expressed as  
\begin{eqnarray}
{\lambda_m}^n &\equiv & (a_i^{g_{\rm b}}){}^n\,({b_i^{g_{\rm b}}})_m 
- (b_i^{g_{\rm b}}){}^n\,({a_i^{g_{\rm b}}})_m\,, \no \\
{\lambda_m}^{np} &\equiv & (a_i^{g_{\rm b}})^{np}\,({b_i^{g_{\rm b}}})_m 
- (b_i^{g_{\rm b}})^{np}\,({a_i^{g_{\rm b}}})_m\,, 
\label{def lambda}
\end{eqnarray}
where $(a_i^{g_{\rm b}})^m$, $(a_i^{g_{\rm b}})^{mn}$, 
$(b_i^{g_{\rm b}})^m$ and $(b_i^{g_{\rm b}})^{mn}$ 
are defined as 
\begin{eqnarray}
a_i^{g_{\rm b}}  &\equiv& g_{\rm b}^{-1}\, a_i\, g_{\rm b} 
= (a_i^{g_{\rm b}})^{m}\,\gP_m+\frac{1}{2}(a_i^{g_{\rm b}})^{mn}\,\gJ_{mn}\,, 
\no \\
b_i^{g_{\rm b}} &\equiv& g_{\rm b}^{-1}\, b_i\, g_{\rm b} 
= (a_i^{g_{\rm b}})^{m}\, \gP_m + \frac{1}{2}(b_i^{g_{\rm b}})^{mn}\,\gJ_{mn}\,.
\end{eqnarray}

Now the action of $\oinv_{(0)}$, $\oinv_{(1)}$ and $\oinv_{(2)}$ can be examined as follows.

\medskip

The action of $\oinv_{(0)}$ is given by 
\begin{eqnarray}
\oinv_{(0)}(\gP_m)&\equiv& {k_m}^n\,\gP_n+\frac{1}{2}{l_m}^{np}\,\gJ_{np}\,, \no \\
\oinv_{(0)}(\gJ_{mn})&=&\gJ_{mn}\,,\qquad \oinv_{(0)}(\gQ^I)=\gQ^I\,,
\end{eqnarray}
where ${k_m}^n$ is determined by the following relation: 
\begin{eqnarray}
{k_m}^n = (\delta-2\eta \lambda)^{-1}{}_m{}^n\,. 
\end{eqnarray}
When $\eta=0$, ${k_m}^n$ is reduced to $\delta_m{}^n$. 
Here we have not displayed the explicit form of ${l_m}^{np}$, 
because it does not appear in the final expression 
due to the presence of the projection operators. 

\medskip 

Then the action of $\oinv_{(1)}$ is written as 
\begin{eqnarray}
\oinv_{(1)}(\gP_m)
&=&i \epsilon^{IJ}k_m{}^n\eta\lambda_n{}^p \,\gQ^J\,\gamma_p\theta_I 
+ \frac{1}{2}\delta^{IJ}k_m{}^n\eta\lambda_n{}^{pq}\,\gQ^J\,\gamma_{pq}\theta_I\,, \no \\ 
\oinv_{(1)} (\gJ_{mn}) &=& 0\,, \no \\
\oinv_{(1)}(\gQ^I) &=& i\sigma_3^{IJ}\,k_m{}^p\eta\lambda^{nm}\,\bar{\theta}_J\gamma_n 
\,\gP_p + \frac{1}{2}\sigma_1^{IJ}\,k_m{}^q\eta\lambda^{m,np}\,\bar{\theta}_J\gamma_{np}
\,\gP_q + \mbox{terms with}~\gJ \,. 
\end{eqnarray}
Here, the terms proportional to $\gJ_{mn}$ are not explicitly written down  
because they do not contribute to the final expression. 

\medskip

Finally, the action of $\oinv_{(2)}$ is evaluated as 
\begin{eqnarray}
\oinv_{(2)}(\gP_m) &=& \bar{\theta}_I\left[
\delta^{IJ}{(\cM_{(2)}^{P\delta})_m}^n
+\epsilon^{IJ}{(\cM _{(2)}^{P\epsilon})_m}^n 
+\sigma_1^{IJ}{(\cM _{(2)}^{P\sigma_1})_m}^n 
+ \sigma_3^{IJ}{(\cM _{(2)}^{P\sigma_3})_m}^n \right]
\theta_J\,\gP_n \no \\
&& + ~\mbox{terms with}~\gJ\,, \no \\
\oinv_{(2)}(\gJ_{mn})&=&0\,,\qquad \oinv_{(2)}(\gQ^I) 
= \mbox{irrelevant terms}\,, 
\end{eqnarray}
where $\cM^{P\delta}_{(2)}$, $\cM^{P\epsilon}_{(2)}$, 
$\cM^{P\sigma_1}_{(2)}$, and $\cM^{P\sigma_3}_{(2)}$ are defined as 
\begin{eqnarray}
{(\cM_{(2)}^{P\delta})_m}^n 
&\equiv&-\frac{i}{4}\,\Bigl[(k_r{}^{n}\ga^r)(k_m{}^s\eta\la_s{}^{pq}\ga_{pq})-(k_r{}^n\eta\la^{r,\,pq}\ga_{pq})(k_m{}^s\ga_s)\Bigr]\,, \no \\
{(\cM _{(2)}^{P\epsilon})_m}^n 
&\equiv& -\frac{1}{2}\Bigl[(k_p{}^n\gamma^p)(k_m{}^s\eta\la_s{}^q\ga_q)-(k_p{}^n\eta\la^{pq}\ga_q)(k_m{}^s\ga_s)\Bigr]\,, \no \\
{(\cM _{(2)}^{P\sigma_1})_m}^n 
&\equiv& (k_s{}^n\eta\la^{rs}\ga_r)
(k_m{}^q\eta\la_q{}^p\ga_p) 
+ \frac{1}{4}(k_s{}^n\eta\la^{s\,,rt}\ga_{rt})
(k_m{}^u\eta\la_u{}^{pq}\ga_{pq})\,, \no \\
{(\cM _{(2)}^{P\sigma_3})_m}^n&\equiv&\frac{i}{2}\,
\Bigl[ ({k_s}^n\eta\la^{rs}\ga_r)(k_m{}^t\eta\la_t{}^{pq}\ga_{pq})
+(k_s{}^n\eta\la^{s\,,rt}\ga_{rt})(k_m{}^q\eta\la_q{}^p\ga_p)\Bigr]\,.
\end{eqnarray}
Here, the terms proportional to $\gJ_{mn}$ have not been written down on the same reasoning. 
Furthermore, the explicit expression of $\oinv_{(2)}(\gQ^I)$ is not necessary for our argument 
because it always leads to higher-order contributions with $\cO(\theta^4)$ in the resulting Lagrangian. 

\medskip 

Next is to evaluate the Lagrangian using the formulae obtained above.

\subsection{The deformed Lagrangian at order $\theta^2$}

Let us now examine the deformed action at the second order of $\theta$. 

\medskip 

Now the Lagrangian in (\ref{YBsM}) can be rewritten as 
\begin{eqnarray}
\mL =-\frac{\sqrt{\la_c}}{4}\left(\gamma^{ab}-\epsilon^{ab}\right)
\str \left[\tilde{d}(A_a)\oinv (A_b)\right]\,, 
\end{eqnarray}
where $\tilde{d}$ is defined as 
\begin{eqnarray}
\tilde{d} \equiv -P_1 + 2P_2 +P_3\,.  
\label{op-tilde-d}
\end{eqnarray}
This Lagrangian can be expanded in terms of $\theta$ as 
\begin{eqnarray}
\mL = \mL\dzero + \mL_{(2,0,0)} + \mL_{(0,0,2)} + \mL_{(1,1,0)} 
+ \mL_{(0,1,1)}  + \mL_{(0,2,0)} + \mL_{(1,0,1)} + \cO(\theta^4)\,. 
\label{expansion}
\end{eqnarray}
Here, $\mL\dzero$ does not include any $\theta$. The second-order term $\mL_{(l,m,n)}$ 
contains two $\theta$s. The set of subscripts $(l,m,n)$ indicates the numbers of $\theta$ included 
in $\tilde{d}(A_a)$, $\oinv$ and $A_b$, respectively.
For example, in the case of $\mL_{(2,0,0)}$, the two $\theta$s are included in $\tilde{d}(A_a)$, 
and there is no $\theta$ in $\oinv$ and $A_b$. 
That is, $\mL_{(2,0,0)}$ is given by 
\begin{eqnarray}
\mL_{(2,0,0)}=-\frac{\sqrt{\la_c}}{4} \left(\gamma^{ab}-\epsilon^{ab}\right) 
\str \left[\tilde{d}((A\dtwo)_a)\oinv\dzero ((A\dzero)_b)\right]\,. \label{200}
\end{eqnarray}

\medskip 

In the following, let us see each term of the expansion (\ref{expansion}). 
The first one is $\mL\dzero$ and does not contain any fermions. 
This can be rewritten into the standard form as follows: 
\begin{eqnarray}
\mL\dzero&=&-\frac{\sqrt{\la_c}}{4} \,(\gamma^{ab}-\epsilon^{ab})\,
\str \left[\tilde{d}((A\dzero)_a)\oinv\dzero ((A\dzero)_b)\right] \no\\
&=&-\frac{\sqrt{\la_c}}{2} \,(\gamma^{ab}-\epsilon^{ab})\, 
e_a^m e_b^n\, k_{nm}\no\\
&=&-\frac{\sqrt{\la_c}}{2}\left[ 
\gamma^{ab} e_\mu^m e_\nu^n \,k_{(mn)} \,
\partial_a X^\mu \partial_b X^\nu
-\epsilon^{ab} e_\mu^m e_\nu^n\, k_{[nm]}\,  
\partial_a X^\mu \partial_b X^\nu \right]\,. \label{last}
\end{eqnarray}
Here we have used the relation 
$e^m_a = e^m_\mu \partial_a X^\mu$, and the $X^\mu$s are the target-spacetime coordinates. 
The last expression (\ref{last}) should be compared with the standard bosonic string action
\begin{eqnarray}
\mL^{\rm b} &=&-\frac{\sqrt{\la_c}}{2}\left[
\gamma^{ab}\, \widetilde{G}_{MN}\, 
\partial_a X^M \partial_b X^N
-\epsilon^{ab} \,B_{MN}\,
\partial_a X^M \partial_b X^N\right]
\end{eqnarray}
with the spacetime metric $\widetilde{G}$ and NS-NS two-form $B$.  
Then one can obtain the following relations: 
\begin{eqnarray}
\widetilde{G}_{MN}&\equiv& e_M^m e_N^n\, k_{(mn)} 
= \tilde{e}^m_M\tilde{e}_{mN}\,, \no \\
B_{MN}&\equiv& e_M^m e_N^n \,k_{[nm]}\,. 
\end{eqnarray}
Here we have introduced the vielbeins $\tilde{e}^m_M$ 
for the deformed metric for our later convenience. 
Note that the index $M$ is raised and lowered 
by $\widetilde{G}^{MN}$ and $\widetilde{G}_{MN}$, 
respectively.

\medskip 

Then let us evaluate the combination $\mL_{(2,0,0)}+\mL_{(0,0,2)}$. 
From the point of view of symmetry, this combination is convenient 
and can be evaluated as 
\begin{eqnarray}
&&\mL_{(2,0,0)}+\mL_{(0,0,2)}\no\\
&=&-\frac{\sqrt{\la_c}}{4}\,(\gamma^{ab}-\epsilon^{ab})\,
\str\bigl[i\bth_I\gamma^m D^{IJ}_a \theta_J\,\gP_m \,e^n_b k_n{}^p\,\gP_p
+ie^m_a \gP_m\,\bth_I\gamma^nD^{IJ}_b \theta_J\, k_n{}^p\gP_p\bigr] \no \\
&=&-\frac{i{\sqrt{\la_c}}}{4}\,(\gamma^{ab}-\epsilon^{ab})\,
\bth_I \left(e^n_b k_{nm}\ga^m D^{IJ}_a
+e^m_a k_{nm}\ga^n D^{IJ}_b \right)\theta_J\,. 
\end{eqnarray}
By the same reasoning, it is helpful to evaluate the combination $\mL_{(1,1,0)}+\mL_{(0,1,1)}$. 
The resulting expression is given by 
\begin{eqnarray}
&&\mL_{(1,1,0)}+\mL_{(0,1,1)} \\
&=&-\frac{\sqrt{\lambda_c}}{4}\,(\gamma^{ab}-\epsilon^{ab})\,
\str\bigl[\sigma_3^{IJ} \gQ^J D^{IK}_a \theta_K\oinv_{(1)}
(e^m_b \gP_m)+ 2e^m_a \gP_m \oinv_{(1)}(-\gQ^ID^{IJ}_b \theta_J)\bigr] \no \\
&=&-\frac{\sqrt{\lambda_c}}{2}\,(\gamma^{ab}-\epsilon^{ab})\, \bth_I 
\left[i\eta\lambda_n{}^p\gamma_p\sigma_3^{IJ}
-\frac{1}{2}\eta\la_n{}^{pq}\gamma_{pq}\sigma_1^{IJ}\right] 
\left(e^m_b k_m{}^nD_a^{JK}+e^m_a k^n{}_mD^{JK}_b\right)\theta_K\,. \no
\end{eqnarray} 
Finally, $\mL_{(0,2,0)}$ and $\mL_{(1,0,1)}$ are evaluated as, respectively,   
\begin{eqnarray}
\mL_{(0,2,0)} &=&-\frac{\sqrt{\lambda_c}}{4}\,(\gamma^{ab}-\epsilon^{ab})\,
\str\left[2e^m_a \gP_m \oinv_{(2)}(e^n_b \gP_n)\right] \no \\
&=& -\frac{\sqrt{\lambda_c}}{2}\,(\gamma^{ab}-\epsilon^{ab})\, e^m_a e^n_b\, \bth_I
\Bigl[\epsilon^{IJ}(\cM _{(2)}^{P\epsilon})_{nm}+\delta^{IJ}(\cM_{(2)}^{P\delta})_{nm} \no \\
&& \qquad\quad + \sigma_1^{IJ}{(\cM _{(2)}^{P\sigma_1})_{nm}} 
+ \sigma_3^{IJ}{(\cM _{(2)}^{P\sigma_3})_{nm}} \Bigr]\theta_J\,, \\
\mL_{(1,0,1)} &=& -\frac{\sqrt{\lambda_c}}{4}\,(\gamma^{ab}-\epsilon^{ab})\,
\str\left[\sigma_3^{IJ}\gQ^J D_a^{IK}\theta_K(-\gQ^LD_b^{LM}) \theta_M \right] \no \\
&=& -\frac{i\sqrt{\lambda_c}}{2}\,\epsilon^{ab}\sigma_3^{IJ}\,
\bth_I e^m_a \gamma_mD_b^{JK} \theta_K\,.
\end{eqnarray}

So far, we have derived the deformed Lagrangian at the quadratic level of $\theta$. 
However, the resulting sum of the components evaluated above is quite intricate and 
we still need to recast it into the canonical form via coordinate transformations.

\subsection{The canonical form of the Lagrangian} 

Here, let us perform coordinate transformations in order to realize the canonical form of the Lagrangian. 
This process is mainly composed of two steps, 1) the shift of $X$ and 2) the rotation of $\theta$. 
%In the following, let us see each of the steps. 

\subsubsection{The canonical form}

First of all, let us present the canonical form of the Lagrangian at order $\theta^2$\,\cite{CLPS}:
\begin{eqnarray}
\mathcal{L}^{(\theta^2)}&=&-\frac{\sqrt{\lambda_c}}{2}\, 
i\bar{\Theta}_I (\gamma^{ab}\delta^{IJ}+\epsilon^{ab}\sigma_3^{IJ})\, 
\tilde{e}_a^m \Gamma_m \widetilde{D}^{JK}_{b}\Theta_K\,, \no \\
\widetilde{D}^{IJ}_{a} &=& \delta^{IJ}\left(\partial_a
-\frac{1}{4}\tilde{\omega}_a^{mn}\Gamma_{mn}\right) 
+\frac{1}{8}\sigma_3^{IJ}\tilde{e}^m_a H_{mnp} \Gamma^{np} \no \\
&& \quad -\frac{1}{8}{\rm e}^{\Phi}\left[\epsilon^{IJ} \Gamma^p F_p 
+ \frac{1}{3!}\sigma_1^{IJ}\Gamma^{pqr}F_{pqr} 
+ \frac{1}{2\cdot 5!}\epsilon^{IJ}\Gamma^{pqrst}F_{pqrst}\right]\tilde{e}^m_a \Gamma_m\,. 
\label{canonical}
\end{eqnarray}
Here, the $\Gamma_m$s are $32\times 32$ gamma matrices composed of  
$\gamma_{\check{m}}$ and $\gamma_{\hat{m}}$ as follows:  
\begin{eqnarray}
\Gamma_{\check{m}}=\sigma_1\otimes \gamma_{\check{m}}\,,\quad 
\Gamma_{\hat{m}}=\sigma_2\otimes \gamma_{\hat{m}}\,.
\end{eqnarray}
Then $\Gamma_{m_1...m_2}$ is defined as 
\[
\Gamma_{m_1..m_n}\equiv \frac{1}{n!}\Gamma_{[m_1}...\Gamma_{m_n]}\,.
\]
Now $\Theta^I~(I=1,2)$ are 32-component Majorana spinors defined as
\begin{eqnarray}
\Theta^I\equiv \left(\begin{array}{c} 1\\ 0\\ \end{array}\right)\otimes \theta^I \,,
\qquad \bar{\Theta}\equiv \Theta^\dag \Gamma^0=\Theta^t \mathcal{C}=\left( ~0~\,,~ 1~\right)\otimes \bth^I \,.
\end{eqnarray}
Here, $\mathcal{C}$ is a charge conjugation matrix defined as
\begin{eqnarray}
\mathcal{C} \equiv i\, \sigma_2\otimes K\otimes K\,.
\end{eqnarray}
The canonical Lagrangian (\ref{canonical}) contains the dilaton $\Phi$, 
the three-form field strength $H_3 = dB_2$ ($B_2$~: NS-NS two-form), 
the one-form field strength $F_1 = d\chi$ ($\chi$~: axion or R-R scalar), 
the three-form field strength $F_3 =dC_2$ ($C_2$~: R-R two-form), 
and the five-form field strength $F_5 = dC_4$ ($C_4$~: R-R four-form). 
Thus, after rewriting the quadratic part of the Lagrangian $\mL$ in (\ref{expansion}) 
into the canonical form, by comparing the resulting form with the canonical form 
(\ref{canonical}), one can read off the component fields of type IIB supergravity. 

\medskip 

The remaining task is to rewrite the Lagrangian $\mL$ expanded above
by performing a shift of $X$ and a rotation of $\theta$\,. We will explain each of the steps below. 

\subsubsection{Shift of $X$}

Let us see the terms with $\gamma^{ab}\partial_{b}\theta$ 
in $\mL$. The relevant parts are 
\[
(a) \quad \mathcal{L}^\gamma_{(2,0,0)}+\mathcal{L}^\gamma_{(0,0,2)} \qquad \mbox{and} \qquad
(b) \quad \mathcal{L}^\gamma_{(1,1,0)}+\mathcal{L}^\gamma_{(0,1,1)}\,. 
\]
One can realize that the terms should appear with $\delta^{IJ}$ 
from the expression of the canonical form (\ref{canonical}). 
There is no obstacle for (a), however (b) involves terms like 
\begin{eqnarray}
\frac{\sqrt{\lambda_c}}{2}\,\bth_I\, \gamma^{ab}\,\sigma_1^{IJ}\,e^m_a\, 
k_{(mn)}\eta\,\lambda^{n\,, pq}\gamma_{pq} \partial_b \theta_J\,. \label{remove}
\end{eqnarray}
Such terms do not appear in the canonical form (\ref{canonical}) and hence must be removed somehow. 
A possible resolution is to shift $X$ as \cite{ABF2}  
\begin{eqnarray}
\label{shiftX}
X^\mu\longrightarrow X^\mu+\bth_I\delta X^{\mu\,IJ}\theta_J\,, \qquad 
\delta X^{\mu\,IJ} \equiv \frac{1}{4}\sigma_1^{IJ}e^{n\,\mu}\,\eta\,\la_n{}^{pq}\gamma_{pq}\,. 
\end{eqnarray}
While this shift removes the problematic terms, it generates additional ones:
\begin{eqnarray}
&& -\frac{\sqrt{\lambda_c}}{2}\,i\bth_I \, \gamma^{ab}\delta^{IJ} 
\Bigl[ \, -\frac{i}{2} \,\sigma_1^{JK}\,e^m_a\, e^n_N\, k_{(mn)} \partial_b 
(e^{n\,N}\,\eta\,\lambda_n{}^{pq}\gamma_{pq} )\no\\
&& \hspace*{3.5cm} -\frac{i}{4}\sigma_1^{JK}\,\partial_P \tilde{G}_{MN} \partial_a X^M \,
\partial_b X^N\,e^{n\,P} \,\eta\,\lambda_n{}^{pq}\gamma_{pq}\Bigr]\theta_K\,. 
\end{eqnarray}
Note here that these terms do not involve derivatives of $\theta$. 

\medskip 

At this stage, the quadratic Lagrangian including $\gamma^{ab}$ is written down as  
\begin{eqnarray}
\mathcal{L}^\gamma&=&-i\frac{\sqrt{\lambda_c}}{2}\,\gamma^{ab}\delta^{IJ}\bth_I 
\Biggl[e^p_a\, k_{(pn)}\, 
(\eta^{nm}-(-1)^J 2 \eta \lambda^{nm} ) \gamma_m 
 D^{JK}_b\no\\
&&+\frac{1}{2}\sigma_3^{JK}\,e^m_a k_{(mn)}\eta\lambda^{n,\,pq}
\gamma_{pq}~e^r_b\,\gamma_r
-\frac{1}{4}\delta^{JK}\,e_a^m\, k_{nm}
\left(\eta^{np}-(-1)^J\, 2\eta\, \lambda^{np}\right)\, 
\gamma_p~ e^q_b k_q{}^r\,\eta\,\lambda_r{}^{st}\,\gamma_{st}\no\\
&&+\frac{1}{4}\delta^{JK}\,e^m_a k_{mn}\, \eta\,\lambda^{n,\,pq}\, 
\gamma_{pq}\, e^r_b \,k_{rs} 
\left(\eta^{st}-(-1)^J\,2\eta\,\lambda^{st}\right)\,\gamma_t\no\\
&&+\frac{i}{2}\epsilon^{JK}e_a^m\, k_{nm} 
\left(\eta^{np}-(-1)^J\,2\eta\,\lambda^{np}\right)\,
\gamma_p~e^q_b\,k_q{}^r\,\eta\,\lambda_r{}^s\,\gamma_s\no\\
&&-\frac{i}{4}\sigma_1^{JK}\,e^m_a k_{(mn)}\,\eta\,\lambda^{n,\,pq}\,
\gamma_{pq}~\omega_b^{rs}\,\gamma_{rs}
-\frac{i}{2}\epsilon^{JK}\,e^m_a\,k_{nm}\,\eta\,\lambda^{np}\,\gamma_p~e^q_b\,k_q{}^r\,
\gamma_r\no\\
&&-\frac{i}{4}\sigma_1^{JK}\,e^m_a\,k_{nm}\,\eta\,\lambda^{n,pq}\,
\gamma_{pq}~e^r_b\,k_r{}^s\,\eta\,\lambda_s{}^{tu}\,\gamma_{tu}-\frac{i}{2}\sigma_1^{JK}\, 
e^m_a\,e^n_M\,k_{(mn)}\,\partial_b\left(e^{p,\,M}\,\eta\,\lambda_p{}^{qr}\,
\gamma_{qr}\right)\no\\
&&-\frac{i}{4}\sigma_1^{JK}\partial_P\widetilde{G}_{MN}\partial_a\,X^M\,
\partial_b\,X^N\,e^{m,\,P}\,\eta\,\lambda_m{}^{np}\,\gamma_{np}
\Biggr]\theta_K\,.
\end{eqnarray}

\medskip 

The next step is to see the terms with $\epsilon^{ab}\partial_{b}\theta$ in $\mL$. 
This part has the terms involving $\sigma_1^{IJ}$ as well. Fortunately, 
the sift of $X$ in (\ref{shiftX}) can eliminate the problematic terms simultaneously, 
while some additional terms including $\epsilon^{ab}$ are again generated. 
Then the quadratic Lagrangian including $\epsilon^{ab}$ is written down as  
\begin{eqnarray}
\mathcal{L}^\epsilon&=&-i\, \frac{\sqrt{\la_c}}{2}\,\epsilon^{ab}\,\bth_I
\Biggl[
\Bigl(
\delta^{IJ}\,e^m_a\,k_{[mn]}\,\gamma^n
+\sigma_3^{IJ}\bigl[e^m_a\,k_{[mn]}\,2\,\eta\,\la^{np}\,\gamma_p+e^m_a\,\gamma_m)\bigr]
\Bigr)
D^{JK}_b\no\\
&&+i\,\sigma_1^{IJ}\,e^m_a\,k_{[mn]}\,\eta\,\la^{n\,,pq}\,\gamma_{pq}
\left(-\frac{1}{4}\delta^{JK}\,\omega_b^{rs}\,\gamma_{rs}
+\frac{i}{2}\epsilon^{JK}e_b^r\,\gamma_r \right)\no\\
&&+\frac{1}{4}\,\delta^{IK}e_a^m\, k_{nm}
\left(\eta^{np}-(-1)^I\, 2\eta\, \lambda^{np}\right)\, 
\gamma_p~ e^q_b k_q{}^r\,\eta\,\lambda_r{}^{st}\,\gamma_{st}\no\\
&&-\frac{1}{4}\delta^{IK}\,e^m_a k_{mn}\, \eta\,\lambda^{n,\,pq}\, 
\gamma_{pq}\, e^r_b \,k_{rs} 
\left(\eta^{st}-(-1)^I\,2\eta\,\lambda^{st}\right)\,\gamma_t\no\\
&&-\frac{i}{2}\epsilon^{IK}e_a^m\, k_{nm} 
\left(\eta^{np}-(-1)^I\,2\eta\,\lambda^{np}\right)\,
\gamma_p~e^q_b\,k_q{}^r\,\eta\,\lambda_r{}^s\,\gamma_s\no\\
&&+\frac{i}{2}\epsilon^{IK}\,e^m_a\,k_{nm}\,\eta\,\lambda^{np}\,\gamma_p~e^q_b\,k_q{}^r\,
\gamma_r+\frac{i}{4}\sigma_1^{IK}\,e^m_a\,k_{nm}\,\eta\,\lambda^{n,pq}\,
\gamma_{pq}~e^r_b\,k_r{}^s\,\eta\,\lambda_s{}^{tu}\,\gamma_{tu}\no\\
&&+\frac{i}{2}\sigma_1^{IK}B_{MN}\partial_a X^M\,\partial_b 
\left( e^{n\,N}\,\eta \,\la_n{}^{pq}\, \right)\gamma_{pq}\no\\
&&+\frac{i}{4}\sigma_1^{IK}\partial_PB_{MN}\partial_aX^M\partial_bX^Ne^{n\,P}
\eta\la_n{}^{pq}\ga_{pq}
\Biggr]\theta_K\,.
\end{eqnarray}

For the next step, it is convenient to switch from the $16 \times 16$ gamma matrices $\gamma$ 
to the $32 \times 32$ ones $\Gamma$, and hence we will work in the $32\times 32$ notation 
in the following. The lift-up rule is summarized in Appendix A,
and it is straightforward to rewrite the Lagrangian.

\subsubsection{Rotation of $\theta$}

After shifting $X$, the resulting derivative terms of $\theta$ take the following form:
\begin{eqnarray}
-\frac{\sqrt{\lambda_c}}{2}\,i\bar{\Theta}_I \,\gamma^{ab}\delta^{IJ}\, 
\tilde{e}_{(I)}{}^m_a \, \Gamma_m \partial_{b}\, \Theta_J\,.
\end{eqnarray}
Here, the vielbeins\footnote{
Note that $\tilde{e}_{(I)}{}^m_a$ satisfy the relation
\[
\tilde{e}_{(I)}{}^m_a \, \tilde{e}_{(I)}{}_{b\, m} = e^m_a e^n_b \,k_{(mn)} 
= \widetilde{G}_{\mu\nu} \partial_a X^\mu \partial_b X^\nu 
\qquad (\text{for}~~I=1\,,2)\,.
\]}
$\tilde{e}_{(I)}{}^m_a$ are defined as 
\begin{eqnarray}
\label{viel-I}
\tilde{e}_{(I)}{}^m_a \equiv e^p_a\, k_{(pn)}\, 
\Bigl[ \eta^{nm}-(-1)^I 2 \eta \lambda^{nm} \Bigr]
\end{eqnarray}
and depend on the index $I$. 
Hence we need to perform a Lorentz transformation for the spinor $\theta$ 
to remove the $I$ dependence. 

\medskip 

The first step is to determine the $I$-independent form of the vielbeins as a reference frame. 
Hereafter, it is fixed by taking $I=1$ in (\ref{viel-I}) as 
\begin{eqnarray}
\label{i-indep}
\tilde{e}^m_a = e^p_a\, k_{(pn)}\, \Bigl[
\eta^{nm}+2 \eta \lambda^{nm} \Bigr]\,.
\end{eqnarray}
Then, by performing a Lorentz transformation for $\theta$, this term can be rewritten as
\begin{eqnarray}
&&\bar{\Theta}_I\,\tilde{e}_{(I)}{}^m_a \,\bar{U}_{(I)}\Gamma_m\, U_{(I)}\partial_b \Theta_I 
+ (\text{the derivative term of $U$})\no\\
&=&\bar{\Theta}_I\,\tilde{e}_{(I)}{}^m_a \,\Lambda_{(I)}{}_m{}^n\Gamma_n\,\partial_b \Theta_I 
+ (\text{the derivative term of $U$})\,.
\end{eqnarray}
Note that the Lorentz transformation performed here depends on the index $I$.

\medskip 

In order to realize the $I$-independent form (\ref{i-indep}), the transformation $\Lambda$ 
should be taken as  
\begin{eqnarray}
\Lambda_{(I)}{}_m{}^n = \bigl[\delta_m{}^p+(-1)^I 2\eta \la_m{}^p\bigr] (\delta-2\eta \la)^{-1}{}_p{}^n\,.
\end{eqnarray}
Then the spinor transformation $U_{(I)}$ and its inverse $\bar{U}_{(I)}$ 
have to be determined through the following relation: 
\begin{eqnarray}
\bar{U}_{(I)}\Gamma_m U_{(I)}=\Lambda_{(I)}{}_m{}^n\Gamma_n\,. \label{rel}
\end{eqnarray}
It seems difficult to present simple formulae of  $U_{(I)}$ and $\bar{U}_{(I)}$ that work 
for an ``arbitrary'' classical $r$-matrix. As a matter of course, given an explicit expression of 
a classical $r$-matrix, these quantities can be computed concretely. 

\medskip 

However, at least for a simple class of classical $r$-matrices, we can propose the following concise forms:  
\begin{eqnarray}
U_{(I)} &=& \frac{1}{\det(\mathbf{1}_{32}+\frac{1}{2}\left[1+(-1)^I\right]\eta \lambda^{mn} 
\Gamma_{mn})^{\frac{1}{32}}} 
\left(\mathbf{1}_{32}+\frac{1}{2} \left[1+(-1)^I\right]\eta \lambda^{mn} \Gamma_{mn} \right)\,, \no\\
\bar{U}_{(I)} &=& \frac{1}{\det(\mathbf{1}_{32}+\frac{1}{2}\left[1+(-1)^I\right]\eta \lambda^{mn} 
\Gamma_{mn})^{\frac{1}{32}}} 
\left(\mathbf{1}_{32}-\frac{1}{2} \left[1+(-1)^I\right]\eta \lambda^{mn} \Gamma_{mn} \right)\,. 
\label{conjecture}
\end{eqnarray}
For example, these are valid for the examples presented in Sect.\ 4. 
It has not been definitely clarified yet to what extent the formulae in (\ref{conjecture}) are valid. 
With our current techniques, if the expressions in (\ref{conjecture}) do not satisfy the relation (\ref{rel}) 
for a given classical $r$-matrix, then it is necessary to derive the concrete forms 
of $U_{(I)}$ and $\bar{U}_{(I)}$ on a case-by-case basis.  

\medskip 

After all this, we have obtained the canonical form of the Lagrangian.
In the actual derivation of R-R fluxes and the dilaton, 
we {\it still} need to use a concrete expression of a classical $r$-matrix and 
computation software like Mathematica or Maple, at least at the current level of understanding.  
We will present the resulting backgrounds for some example classical $r$-matrices in Sect.\ 4.

\subsection{The master formula for the dilaton} 

We propose the master formula for dilaton, 
in which the dilaton is described in terms of the classical $r$-matrix directly. 
That is, just by putting the elements of the classical $r$-matrix, the associated dilaton is 
obtained directly, without passing through the supercoset construction.
The formula is given by 
\begin{eqnarray}
{\rm e}^{\Phi} &=& \frac{1}{\det_{32} (\mathbf{1}_{32}+\eta\, \lambda^{mn} 
\Gamma_{mn})^{\frac{1}{32}}} \nonumber \\ 
&=& \frac{1}{\det_{10} (\mathbf{\delta}_{m}^{~~n}+2\,\eta\, \lambda_{m}^{~~n})^{\frac{1}{2}}}\,, 
\label{master}
\end{eqnarray}
where $\det_{D}$ means the determinant of a $D \times D$ matrix. 
Recall that $\lambda_{m}^{~~n}$, which is defined in (\ref{def lambda}), is determined 
by putting the elements of the classical $r$-matrix.
Although this formula has not been proven and just a conjectured form, 
it works well for well-known examples, including the examples discussed in Sect.\ 4.

\medskip 

Similar master formulae may be derived for other R-R fluxes, 
though we have not succeeded in deriving them yet. 
It is important to try to complete the master formulae and 
directly check the on-shell condition of type IIB supergravity.

\section{Examples}

Let us consider some examples of classical $r$-matrices satisfying the homogeneous CYBE. 
Then it is possible to complete the supercoset construction and derive the resulting backgrounds. 

\medskip

For the following argument, let us introduce the terms ``abelian'' and ``non-abelian'' 
classical $r$-matrices. Suppose that a classical $r$-matrix is given by $r = a\wedge b$. 
It is called ``abelian'' when $a$ and $b$ commute with each other.
If not, it is ``non-abelian''.

\subsection{Gravity duals of noncommutative gauge theories}

Let us discuss gravity duals of noncommutative gauge theories as Yang-Baxter deformations 
with the following classical $r$-matrix \cite{MR-MY}: 
\begin{eqnarray}
r = P_{2} \wedge P_{3}\,. 
\label{r-MR}
\end{eqnarray}
Here it is assumed that  $P_{\mu}$ are naturally embedded into $8\times 8$ matrices like 
\begin{eqnarray}
\label{embedding-88}
\begin{pmatrix}
\;~P_\mu&~\boldsymbol{0_4}~\\
~\boldsymbol{0_4}&\boldsymbol{0_4}
\end{pmatrix}\,.
\end{eqnarray}
This is an abelian classical $r$-matrix and satisfies the homogeneous CYBE. 

\medskip 

The bosonic part has already been studied in\cite{MR-MY}, where the string-frame metric 
and NS-NS two-form are reproduced with the $r$-matrix (\ref{r-MR}). 
The R-R sector and dilaton can be determined by performing the supercoset construction. 

\medskip 

The supercoset construction can be carried out by following the general argument in Sect.\ 3.
In the present case, the key ingredient $\lambda_{mn}$ is given by
\begin{eqnarray}
\lambda_{mn}=\begin{pmatrix}
\;\lambda_{\check{m}\check{n}}~&~\boldsymbol{0_5}~\\
\boldsymbol{0_5}&\boldsymbol{0_5}
\end{pmatrix}\,,\quad
 \lambda_{\check{m}\check{n}}=
 \begin{pmatrix}
\;~0~&~0~&~0~&~0~&~0~\\
~0~&~0~&~0~&~0~&~0~\\
~0~&~0~&~0~&~-z^{-2}~&~0~\\
~0~&~0~&~z^{-2}~&~0~&~0~\\
~0~&~0~&~0~&~0~&~0~
\end{pmatrix}
\,. 
\label{lambda-MR}
\end{eqnarray}
Then $U_{(I)}$ and $\bar{U}_{(I)}$ are obtained by using the formulae in (\ref{conjecture}).

\medskip 

With the general argument in Sect.\ 3, one can read off the following background: 
\begin{eqnarray}
ds^2 &=&  \frac{-(dx^0)^2 + (dx^1)^2}{z^2} + \frac{z^2\left[
(dx^2)^2 + (dx^3)^2
\right]}{z^4 +4\eta^2}
+\frac{dz^2}{z^2} + ds^2_{\rm S^5}\,, \nonumber \\ 
B_2 &=& \frac{2\eta}{z^4+4\eta^2}\,dx^2 \wedge dx^3\,, \nonumber \\ 
F_3 &=&\frac{8\,\eta}{z^5}\, dx^0 \wedge dx^1 \wedge dz\,, \nonumber \\ 
F_5 &=& 4 \left(\text{e}^{2 \Phi}\,\omega_{\text{AdS}_5}
+\omega_{\text{S}^5}\right) \,, \qquad \Phi = \frac{1}{2}\log \left(\frac{z^4}{z^4+4\eta^2}\right)\,. 
\end{eqnarray}
This is nothing but the solution found in \cite{HI,MR} as a gravity dual of noncommutative gauge theories. 
Note that the dilaton can be reproduced by using the master formula 
(\ref{master}) with (\ref{lambda-MR}).

\subsection{$\gamma$-deformations of S$^5$}

We shall discuss three-parameter $\gamma$-deformations of S$^5$ 
with the following classical $r$-matrix \cite{LM-MY}: 
\begin{eqnarray}
r = \frac{1}{8} \left(\nu_3\, h_1 \wedge h_2 + \nu_1\, h_2 \wedge h_3 + \nu _2\, h_3 \wedge h_1 \right)\,. 
\end{eqnarray}
Here  $\nu_i~(i=1,2,3)$ are real constant parameters, and $h_a~(a=1,2,3)$ 
are the Cartan generators of $\mathfrak{su}(4)$ embedded in $8\times 8$ matrices 
as the lower diagonal block (for their matrix representation, 
see Appendix A). This is an abelian classical $r$-matrix and satisfies the CYBE. 

\medskip 

The bosonic part has already been studied in \cite{LM-MY}. 
The remaining task is to perform supercoset construction 
in order to determine the R-R sector and dilaton. 

\medskip 

The quantities $U_{(I)}$ and $\bar{U}_{(I)}$ are determined by  
\begin{eqnarray}
\lambda_{mn}=\begin{pmatrix}
\;\boldsymbol{0_5}~&~\boldsymbol{0_5}~\\
\boldsymbol{0_5}&\lambda_{\hat{m}\hat{n}}
\end{pmatrix}\,,\quad
 \lambda_{\hat{m}\hat{n}}=
 \begin{pmatrix}
\;~0~&~0~&-\frac{1}{2}\nu_3\,\rho_1\,\rho_2~&~0~&\frac{1}{2}\nu_1\,\rho_2\,\rho_3~\\
~0~&~0~&~0~&~0~&~0~\\
~\frac{1}{2}\nu_3\,\rho_1\,\rho_2~&~0~&~0~&~0~&-\frac{1}{2}\nu_2\, \rho_3\,\rho_1~\\
~0~&~0~&~0~&~0~&~0~\\
~-\frac{1}{2}\nu_1\,\rho_2\,\rho_3~&~0~&\frac{1}{2}\nu_2\, \rho_3\,\rho_1~&~0~&~0~
\end{pmatrix}
\,. 
\label{lambda-LM}
\end{eqnarray}

\medskip

Then, by following the general discussion, 
the full solution presented in \cite{LM,Frolov} can be reproduced as 
\begin{eqnarray}
ds^2 &=& ds^2_{\rm AdS_5} + \sum_{i=1}^3(d\rho_i^2+G \,\rho_i^2d\phi_i^2) 
+ G \rho_1^2\rho_2^2\rho_3^2 \left(\sum_{i=1}^3\hat{\gamma}_i\, d\phi_i\right)^2\,,   
\label{3-metric} \\ 
B_2 &=& G \left(
\hat{\gamma}_3\,\rho_1^2\rho_2^2\,d\phi_1\wedge d\phi_2 
+ \hat{\gamma}_1\,\rho_2^2\rho_3^2\,d\phi_2\wedge d\phi_3  
+ \hat{\gamma}_2\,\rho_3^2\rho_1^2\,d\phi_3\wedge d\phi_1 
\right)\,, \no \\ 
F_3 &=& -4 \sin^3\alpha\,\cos \alpha\, \sin \theta\, \cos \theta \,
\left(\sum_{i=1}^3\hat{\gamma}_i\,\,d\phi_i\right)\wedge d\alpha \wedge d\theta  \nonumber \\ 
F_5 &=&4\left(\omega_{\text{AdS}_5}+G\, \omega_{\text{S}^5}\right) \,, \qquad 
\Phi = \frac{1}{2}\log\,G \,.
\label{3-NSNS}
\end{eqnarray}
Here we have introduced  a scalar function $G$ and $\hat{\gamma}_i~(i=1,2,3)$ defined as  
\begin{eqnarray}
G^{-1} \equiv 1 + \hat{\gamma}_3^2\,\rho_1^2\rho_2^2 + \hat{\gamma}_1^2\,\rho_2^2\rho_3^2 
+ \hat{\gamma}_2^2\,\rho_3^2\rho_1^2\,, 
\qquad \hat{\gamma}_i\equiv \eta \nu_i\,.
\end{eqnarray}
Three coordinates $\rho_i$ satisfying the constraint $\sum_{i=1}^3\rho_i^2 =1$ 
are parametrized by two angle variables $\alpha$ and $\theta$ through the relation:
\begin{eqnarray}
\rho_1\equiv \sin \alpha \cos \theta\,,\quad \rho_2\equiv 
\sin \alpha \sin \theta\,,\quad \rho_3\equiv \cos \alpha\,. 
\end{eqnarray}

\medskip 

It should be remarked that the resulting background is non-supersymmetric 
other than for exceptional cases like $\nu_1=\nu_2=\nu_3$. 
But the supercoset construction still works well. Note that the dilaton can be reproduced 
by using the master formula (\ref{master}) with (\ref{lambda-LM}).

\subsection{Schr\"odinger spacetimes}

Let us consider Schr\"odinger spacetimes by employing the following 
classical $r$-matrix \cite{Sch-MY}: 
\begin{eqnarray}
r = \frac{i}{4}\,P_- \wedge (h_1 + h_2 + h_3)\,. \label{r-Sch}
\end{eqnarray}
Here $P_- \equiv (P_0 - P_3)/\sqrt{2}$ is a light-cone generator in $\mathfrak{su}(2,2)$, 
and $h_1,h_2,h_3$ are the Cartan generators in $\mathfrak{su}(4)$\,. 

\medskip

Note here that the classical $r$-matrix (\ref{r-Sch}) contains a tensor product 
of an $\mathfrak{su}(2,2)$ generator and an $\mathfrak{su}(4)$ one. 
Hence the rotation of $\theta$ should be a ten-dimensional Lorentz transformation,
%10D Lorentz transformation in essential 
and becomes intricate. The quantities $U_{(I)}$ and $\bar{U}_{(I)}$ are given by, 
respectively, 
\begin{eqnarray}
\lambda_{mn} = \begin{pmatrix}
\;\boldsymbol{0_5}~&~\lambda_{\check{m}\hat{n}}~\\
\lambda_{\hat{m}\check{n}}&\boldsymbol{0_5}
\end{pmatrix}\,,\quad
 \lambda_{\check{m}\hat{n}}=
 \begin{pmatrix}
\;\frac{\sin r\sin\zeta}{2\sqrt{2}z}~&~0~&\frac{\sin r\cos\zeta}{2\sqrt{2}z}~&~0~
&\frac{\cos r}{2\sqrt{2}z}~\\
~0~&~0~&~0~&~0~&~0~\\
~0~&~0~&~0~&~0~&~0~\\
\frac{\sin r\sin\zeta}{2\sqrt{2}z}~&~0~&\frac{\sin r\cos\zeta}{2\sqrt{2}z}~&~0~
&\frac{\cos r}{2\sqrt{2}z}~\\
~0~&~0~&~0~&~0~&~0~
\end{pmatrix}
\,. 
\label{lambda-Sch}
\end{eqnarray}
Note that $\lambda_{\hat{m}\check{n}}$ can be obtained from $\lambda_{\check{m}\hat{n}}$, 
because $\lambda_{mn}$ is anti-symmetric.
 
\medskip 

After all, the full solution \cite{MMT} has been reproduced as 
\begin{eqnarray}
ds^2 &=& \frac{-2dx^+dx^- + (dx^1)^2 + (dx^2)^2 +dz^2}{z^2} - \eta^2 \frac{(dx^+)^2}{z^4} 
+ ds^2_{\rm S^5}\,,  \nonumber \\ 
B_2 &=& \frac{\eta}{z^2}\,dx^+ \wedge (d\chi + \omega)\,, \nonumber \\ 
F_5 &=& 4 \left(\omega_{\text{AdS}_5}
+\omega_{\text{S}^5}\right) \,, \qquad  \Phi = \mbox{const.}, \label{Sch}
\end{eqnarray}
and the other fields are zero. 
Here, the light-cone coordinates are defined as 
\[
x^\pm \equiv \frac{1}{\sqrt{2}} (x^0\pm x^3)\,.
\]
The S$^5$ metric is given by 
\begin{align} 
ds^2_{\rm S^5}&=(d\chi+\omega)^2 +ds^2_{\rm \mathbb{C}P^2}\,, \nln 
ds^2_{\rm \mathbb{C}P^2}&= d\mu^2+\sin^2\mu\,
\bigl(\Sigma_1^2+\Sigma_2^2+\cos^2\mu\,\Sigma_3^2\bigr)\,.  
\label{S1overCP2}
\end{align}
Namely, the round S$^5$ is expressed as an S$^1$-fibration over $\mathbb{C}$P$^2$,  
where $\chi$ is the fiber coordinate and $\omega$ is 
a one-form potential of the K\"ahler form on $\mathbb{C}$P$^2$. 
The symbols $\Sigma_i ~(i=1,2,3)$ and $\omega$ are defined as  
\begin{align}
\Sigma_1&= \tfrac{1}{2}(\cos\psi\, d\theta +\sin\psi\sin\theta\, d\phi)\,, \nln 
\Sigma_2&= \tfrac{1}{2}(\sin\psi\, d\theta -\cos\psi\sin\theta\, d\phi)\,, \nln 
\Sigma_3&= \tfrac{1}{2}(d\psi +\cos\theta\, d\phi)\,, 
\qquad 
\omega=\sin^2\mu\, \Sigma_3\,. 
\end{align} 

\medskip 

It should be remarked that the R-R sector has not been deformed and the dilaton remains constant, 
though the expression of the fermionic sector is very complicated in the middle of the computation.  
The cancellation of the deformation effect is really non-trivial. 
Note also that the background (\ref{Sch}) does not preserve any supersymmetries \cite{MMT}. 
It may sound surprising that the supercoset construction still works well 
without the help of supersymmetries.  

\medskip 

The constant dilaton of this background can be reproduced by using the master formula (\ref{master}) 
with (\ref{lambda-Sch}) as well.

\subsection{A non-abelian classical $r$-matrix}

So far, we have considered abelian classical $r$-matrices, for which 
it seems likely that supercoset construction works well 
even though the resulting background is non-supersymmetric. 
The next significant issue is to study non-abelian classical $r$-matrices. 

\medskip 

As for non-abelian classical $r$-matrices, there is no well-known example of the associated background. 
A nice candidate for non-abelian classical $r$-matrices is given by
\begin{eqnarray}
r&=& \frac{1}{\sqrt{2}}E_{24}\wedge (c_1E_{22}-c_2 E_{44}) \qquad \qquad 
\Bigl[(E_{ij}){}_{kl}\equiv \delta_{ik}\,\delta_{jl}\Bigr] \no\\
&=& -\frac{1}{2} P_-\wedge \left[\frac{c_1+c_2}{2}\,\left(D-L_{03}\right)
+i\,\frac{c_1-c_2}{2}\,\left(L_{12}-\frac{i}{2}\,\mathbf{1}_4\right)\right]\,. 
\label{na-r}
\end{eqnarray}
Note here that $\mathbf{1}_4$ is included in the expression and 
hence the image is extended from $\mathfrak{su}(2,2|4)$ 
to $\mathfrak{gl}(4|4)$. However, it can be ignored due to the presence of 
the projection operator in the classical action as pointed out in \cite{Stijn1}. 

\medskip 

To ensure that the resulting metric and NS-NS two-form are real, 
it is necessary to impose the reality condition \cite{MY-duality} 
\begin{eqnarray}
c_2=c_1^\ast\,.
\end{eqnarray}
It is now convenient to introduce $a_1$, $a_2$ as follows: 
\begin{eqnarray}
a_1\equiv\frac{c_1+c_2}{2}=\text{Re}(c_1)\,,\quad 
a_2\equiv \,i\frac{c_1-c_2}{2}= -\text{Im}(c_1)\,.
\end{eqnarray} 
Note here that the classical $r$-matrix (\ref{na-r}) is non-abelian in general. 
The case that $c_1$ is pure imaginary (i.e., $a_1=0$) is exceptional and it becomes abelian. 

\medskip

The bosonic part has already been studied well \cite{KMY-SUGRA,MY-duality,Stijn1}, 
and the remaining task is to determine the R-R sector and dilaton 
by performing supercoset construction. 
The quantities $U_{(I)}$ and $\bar{U}_{(I)}$ are obtained by using 
\begin{eqnarray}
\lambda_{mn} &=& \begin{pmatrix}
\;\lambda_{\check{m}\check{n}}~&~\boldsymbol{0_5}~\\
\boldsymbol{0_5}&\boldsymbol{0_5}
\end{pmatrix}\,,\no\\
 \lambda_{\check{m}\check{n}}&=&
 \frac{1}{2\sqrt{2}}
 \begin{pmatrix}
\;0~&-\frac{a_1\,x^1+a_2\,x^2}{z^2}~&\frac{a_2\,x^1-a_1\,x^2}{z^2}&~0~&-\frac{a_1}{z}~\\
\frac{a_1\,x^1+a_2\,x^2}{z^2}~&~0~&~0~&\frac{a_1\,x^1+a_2\,x^2}{z^2}~&~0~\\
\frac{a_1\,x^2-a_2\,x^1}{z^2}~&~0~&~0~&\frac{a_1\,x^2-a_2\,x^1}{x^2}~&~0~\\
~0~&-\frac{a_1\,x^1+a_2\,x^2}{x^2}~&\frac{a_2\,x^1-a_1\,x^2}{z^2}~&~0~&-\frac{a_1}{z}~\\
\frac{a_1}{z}~&~0~&~0~&\frac{a_1}{z}~&~0~
\end{pmatrix}
\,. 
\label{lambda-HRR}
\end{eqnarray}

\medskip 

After all this, one can read off the resulting background:\footnote{As another possibility, 
one may take a non-constant dilaton $\Phi = \log z$ so as to respect the Bianchi identity $dF_3=0$. 
But in this case the dilaton does not satisfy the equations of motion for the dilaton 
as well as for other components.} 
\begin{eqnarray}
ds^2&=&\frac{-2dx^+dx^-+d\rho^2+\rho^2 d\phi^2+dz^2}{z^2}
- \eta^2 \left[(a_1^2+a_2^2)\frac{\rho^2}{z^6}+\frac{a_1^2}{z^4}\right](dx^+)^2+ds^2_{\text{S}^5}\,, \no\\
B_2 &=& \eta\left[\frac{a_1x^1+a_2x^2}{z^4}dx^+\wedge dx^1
+\frac{a_1x^2-a_2x^1}{z^4}dx^+\wedge dx^2+a_1\frac{1}{z^3}dx^+\wedge dz\right]\,, \no\\
F_3 &=& 4\eta\left[\frac{a_2x^1-a_1x^2}{z^5}dx^+\wedge dx^1\wedge dz 
+\frac{a_1x^1+a_2x^2}{z^5}dx^+\wedge dx^2\wedge dz
+\frac{a_1}{z^4}dx^+\wedge dx^1\wedge dx^2\right]\,, \no\\
F_5 &=& 4 \left(\omega_{\text{AdS}_5}
+\omega_{\text{S}^5}\right) \,,\quad \quad \Phi=\text{const}\,., \label{new}
\end{eqnarray}
and the other components are zero. 
Here the light-cone coordinates are defined in the same way as the previous section.
Notice that the background (\ref{new}) does not satisfy the equation of motion of $B_2$  
because the Bianchi identity for $F_3$ is broken, namely 
\[
dF_3 = 16 \eta\, \frac{a_1}{z^5}\,dx^+\wedge dx^1 \wedge dx^2 \wedge dz  \neq 0\,. 
\]
Thus the classical $r$-matrix (\ref{na-r}) does not lead to a solution of type IIB supergravity. 

\medskip

It is worth noting that the pathology vanishes when $a_1=0$. This is an exceptional case 
in which the classical $r$-matrix becomes abelian and the background (\ref{new})
is reduced to the Hubeny-Rangamani-Ross solution \cite{HRR}. 
This correspondence was originally argued in \cite{MY-duality} and elaborated in \cite{Stijn1}.  
Note here that the constant dilaton can be reproduced by using the master formula (\ref{master}) with 
(\ref{lambda-HRR}) again. 

\medskip 

It would be valuable to see that the background (\ref{new}) is different from 
the one constructed in \cite{KMY-SUGRA}. The former (\ref{new}) does not contain 
the R-R fluxes with the S$^5$ indices, while the latter does. It was conjectured 
in \cite{KMY-SUGRA,MY-duality} that the classical $r$-matrix (\ref{na-r}) should be 
associated with the latter, but it was not correct.  
Our supercoset construction has revealed that the classical $r$-matrix (\ref{na-r}) 
should be associated with the background (\ref{new}). 

\medskip 

The result that the Bianchi identity is broken is similar to 
the $q$-deformed AdS$_5\times$S$^5$ \cite{ABF2}. 
In fact, the background (\ref{new}) satisfies the generalized type IIB supergravity 
equations of motion proposed in \cite{scale}. 
Detailed analysis will be presented in a future paper \cite{future}.

\section{Conclusion and discussion}

In this paper, we have discussed the supercoset construction in the Yang-Baxter deformed 
AdS$_5\times$S$^5$ superstring based on the homogeneous CYBE. 
We have made a general argument without relying on specific expressions of classical $r$-matrices. 
In particular, we have presented the master formula to describe the dilaton 
in terms of classical $r$-matrices. 
The ultimate goal is to represent all of the R-R fluxes as well, and this is a
really fascinating future problem. 
If it is carried out, the on-shell condition of type IIB supergravity 
can be checked directly for general classical $r$-matrices 
and one can test the conjecture of the gravity/CYBE correspondence. 

\medskip 

Then we have explicitly performed supercoset construction for some classical $r$-matrices 
satisfying the homogeneous CYBE. For abelian classical $r$-matrices, 
perfect agreement has been shown for well-known examples including 
gravity duals of non-commutative gauge theories, $\gamma$-deformations of S$^5$ 
and Schr\"odinger spacetimes. Remarkably, the supercoset construction works well, 
even if the resulting backgrounds are not maximally supersymmetric. 
In particular, three-parameter $\gamma$-deformations of S$^5$ and 
Schr\"odinger spacetimes do not preserve any supersymmetries. 
For non-abelian $r$-matrices, we have concentrated on a specific example. 
The resulting background does not satisfy the equation of motion of the NS-NS two-form  
because the Ramond-Ramond three-form is not closed. Thus, at least so far, 
it seems likely that there would be no problem for abelian classical $r$-matrices, 
while there are some potential problems in the non-abelian cases. 
We will report on results on other examples of abelian and 
non-abelian classical $r$-matrices in the near future \cite{future}. 

\medskip 

There are many open problems. The Yang-Baxter deformation has diverse applicability. 
For example, it can be applied to the AdS$_5\times T^{1,1}$ background \cite{CMY}. 
In this case, the Green-Schwarz string action has not yet been constructed.
However, at least for the bosonic sector\footnote{The coset structure 
of $T^{1,1}$ is a little intricate, so even the analysis on the undeformed $T^{1,1}$ is not trivial. }, 
it has been shown that three-parameter 
$\gamma$-deformations of $T^{1,1}$ \cite{LM,CO} can be reproduced as 
Yang-Baxter deformations with abelian classical $r$-matrices \cite{CMY}. It is remarkable 
that the AdS$_5\times T^{1,1}$ background is not integrable because chaotic string solutions exist 
\cite{BZ}. It would be of significance to construct the AdS$_5\times T^{1,1}$ superstring action 
and then investigate its Yang-Baxter deformations by following the procedure presented here.  

\medskip 

It is also interesting to consider a supersymmetric extension of 
Yang-Baxter deformations of Minkowski spacetime 
\cite{YB-Min}. As a toy model along this direction, 
it is easier to study the Nappi-Witten model \cite{NW}. 
Yang-Baxter invariance of this model has been discussed in \cite{KY-NW}. 
It would be nice to argue its supersymmetric extension 
by employing the work \cite{super-NW} 
and further generalization with general symmetric two-forms \cite{SYY}. 

\medskip 

We believe that our supercoset construction could capture the tip of an iceberg, 
namely the gravity/CYBE correspondence that denotes a non-trivial relation 
between type IIB supergravity and the classical Yang-Baxter equation.

\subsection*{Acknowledgments}

We are very grateful to Andrzej Borowiec and Jerzy Lukierski for useful discussions,  
and to Takuya Matsumoto and Jun-ichi Sakamoto for collaborations at an early stage.  
The work of H.K.\ is supported by the Japan Society for the Promotion of Science (JSPS).
The work of K.Y.\ is supported by the Supporting Program for Interaction-based Initiative Team Studies 
(SPIRITS) from Kyoto University and by a JSPS Grant-in-Aid for Scientific Research (C) No.\,15K05051.
This work is also supported in part by the JSPS Japan--Russia Research Cooperative Program 
and the JSPS Japan--Hungary Research Cooperative Program.

\appendix 

\section*{Appendix}

\section{A matrix representation of $\mathfrak{su}(2,2|4)$}

We present here a matrix representation of the superalgebra $\alg{su}(2,2|4)$. 
Our notation and conventions basically follows those utilized in \cite{ABF2}.

\subsection*{A representation of $\alg{su}(2,2)$}

It is convenient to introduce the following basis of $\alg{su}(2,2)\simeq\alg{so}(2,4)$: 
\begin{eqnarray}
\alg{su}(2,2)=\text{span}_{\mathbb{R}}
\left\{~\ga_\mu\,,\ga_5\,,n_{\mu\nu}=\frac{1}{4}[\ga_\mu\,,\ga_\nu]\,, 
n_{\mu5}=\frac{1}{4}[\ga_\mu\,,\ga_5]~|
~~\mu,\nu=0,1,2,3~\right\}\,.
\end{eqnarray}
Here, $\gamma_{\mu}$ are gamma matrices satisfying the Dirac algebra:
\begin{eqnarray}
\{\gamma_{\mu}\,, \gamma_{\nu}\} = 2\eta_{\mu\nu}\,, 
\end{eqnarray}
where $\eta_{\mu\nu}$ is the four-dimensional Minkowski metric with mostly plus. 
It is convenient to adopt the following matrix realizations of the $\gamma_{\mu}$:
\begin{eqnarray}
&& \gamma_1=
\begin{pmatrix}
\;0~&~0~&~0~&-1\\
0&0&1&~0\\
0&1&0&~0\\
-1&0&0&~0\\
\end{pmatrix}\,, 
\quad 
\gamma_2=
\begin{pmatrix}
\;0~&~0~&~0~&~i\\
0&0&i&~0\\
0&-i&0&~0\\
-i&0&0&~0\\
\end{pmatrix}\,, \quad 
\gamma_3=
\begin{pmatrix}
\;0~&~0~&~1~&~0\\
0&0&0&~1\\
1&0&0&~0\\
0&1&0&~0\\
\end{pmatrix}\,,\nonumber \\ 
&& \gamma_0=
%i
-i\gamma_4=
\begin{pmatrix}
\;0~&~0~&1~&0\\
0&0&0&-1\\
-1&0&0&~0\\
0&1&0&~0\\
\end{pmatrix}\,, \quad 
\gamma_5=i\gamma_1\gamma_2\gamma_3\gamma_0=
\begin{pmatrix}
\;1~&~0~&~0~&0\\
0&1&0&~0\\
0&0&-1&~0\\
0&0&0&-1\\
\end{pmatrix}\,.
\end{eqnarray}

\subsubsection*{A conformal basis}

It is also helpful to use the conformal basis: 
\begin{eqnarray}
\mathfrak{so}(2,4) = {\rm span}_{\mathbb{R}}\{~P_{\mu}\,, L_{\mu\nu}\,, D\,, K_{\mu}~|
~~\mu,\nu=0,1,2,3~\}\,.
\label{cb}
\end{eqnarray}
Here, the translation generators $P_{\mu}$, 
the Lorentz rotation generators $L_{\mu\nu}$, the dilatation $D$,   
and the special conformal generators $K_{\mu}$ 
are represented by, respectively,  
\begin{align}
P_{\mu} \equiv \frac{1}{2}(\ga_{\mu}-2n_{\mu5})\,,  \quad 
L_{\mu\nu} \equiv n_{\mu\nu}\,, \quad 
D \equiv \frac{1}{2}\gamma_5\,,  
\quad 
K_{\mu} \equiv \frac{1}{2}(\ga_{\mu}+2n_{\mu5})\,. 
\end{align}
The non-vanishing commutation relations are given by 
\begin{eqnarray}
[P_\mu ,K_\nu]&=& 2 (L_{\mu\nu}+\eta_{\mu\nu}\, D\,)\,,\quad 
[D,P_{\mu}]=P_\mu\,,\quad [D,K_\mu]=-K_\mu\,,\nonumber\\
\left[P_\mu,L_{\nu\rho}\right] &=&\eta_{\mu\nu}\, P_\rho-\eta_{\mu\rho}\, P_\nu \,,\quad 
[K_\mu,L_{\nu\rho}]=\eta_{\mu\nu}\,K_\rho-\eta_{\mu\rho}\,K_\nu\,, \nonumber\\
\left[L_{\mu\nu} ,L_{\rho\sigma}\right]&=&\eta_{\mu\sigma}\, L_{\nu\rho}
+\eta_{\nu\rho}\,L_{\mu\sigma}-\eta_{\mu\rho}\,L_{\nu\sigma}-\eta_{\nu\sigma}\,L_{\mu\rho}\,.
\end{eqnarray}

\subsection*{A representation of $\mathfrak{su}(4)$}

It is easy to see that
\begin{eqnarray}
n_{ij}=\frac{1}{4}\left[\gamma_i,\gamma_j\right] \qquad (i,j=1,\ldots,5)  
\end{eqnarray}
generate $\mathfrak{so}(5)$ by using the Clifford algebra 
\begin{eqnarray}
\left\{\gamma_i,\gamma_j\right\}=2\delta_{ij}\,. 
\end{eqnarray}
Note that $\mathfrak{so}(6)$ is spanned by the set of the Hermite generators,
\begin{eqnarray}
\mathfrak{so}(6) = {\rm span}_{\mathbb{R}}\Bigl\{~\frac{1}{2}\gamma_i\,,~ i\,n_{ij}~\Bigr\}\,.
\end{eqnarray}
It is convenient to introduce the Cartan generators of $\mathfrak{su}(4)$ as follows
\begin{align}
h_1 \equiv 2i\,n_{12}\,,\qquad h_2 \equiv 2i\,n_{43}\,,\qquad h_3 \equiv \gamma_5\,.
\end{align}

\subsection*{An $8 \times 8$ supermatrix representation}

By using the gamma matrices introduced above, 
let us represent the $\mathfrak{su}(2,2|4)$ generators by $8\times 8$ supermatrices. 

\medskip

It is helpful to introduce the following quantities: 
\begin{eqnarray}
\boldsymbol{\gamma}_{\check{m}} 
\equiv \left\{\gamma_0\,,\gamma_1\,,\gamma_2\,,\gamma_3\,,\gamma_5\,\right\}\,, \qquad 
\boldsymbol{\gamma}_{\hat{m}}
\equiv \left\{-\gamma_4\,,-\gamma_1\,,-\gamma_2\,,-\gamma_3\,,-\gamma_5\,\right\}\,. 
\end{eqnarray}
For later convenience, we introduce the following metrics: 
\begin{eqnarray}
\eta_{\check{m}\check{n}} &=& \mbox{diag}(-1,1,1,1,1)\,,\qquad 
\eta_{\hat{m}\hat{n}} = \mbox{diag}(1,1,1,1,1)\,,\no\\
\no\\
\eta_{mn} &=& \begin{pmatrix}
~\eta_{\check{m}\check{n}} & 0 \\ 
0 & \eta_{\hat{m}\hat{n}}~ 
\end{pmatrix}\,. 
\end{eqnarray}

\medskip

Then the $\mathfrak{su}(2,2|4)$ generators $(\gP_{\check{m}}\,, \gP_{\hat{m}}\,, 
\gJ_{\check{m}\check{n}}\,, \gJ_{\hat{m}\hat{n}}\,, \gQ^{I})~(I=1,2)$ 
can be represented by the following $8\times 8$ supermatrices:
\begin{eqnarray}
&& \gP_{\check{m}} = 
\begin{pmatrix}
-\frac{1}{2}\boldsymbol{\gamma}_{\check{m}} & ~~\boldsymbol{0_4}~ \\  
\boldsymbol{0_4} & ~~\boldsymbol{0_4}~ 
\end{pmatrix}
\,, \qquad 
\gJ_{\check{m}\check{n}} = 
\begin{pmatrix}
+\frac{1}{2}\boldsymbol{\gamma}_{\check{m}\check{n}} & ~~\boldsymbol{0_4}~ \\  
\boldsymbol{0_4} & ~~\boldsymbol{0_4}~ 
\end{pmatrix}
 \qquad (\check{m},~\check{n}=0,\ldots,4)\,, \no \\ 
&& \gP_{\hat{m}} = 
\begin{pmatrix}
~\boldsymbol{0_4} & ~\boldsymbol{0_4} \\  
~\boldsymbol{0_4} & ~+\frac{i}{2} \boldsymbol{\gamma}_{\hat{m}} 
\end{pmatrix}
\,, \qquad 
\gJ_{\hat{m}\hat{n}} = 
\begin{pmatrix}
~\boldsymbol{0_4} & ~\boldsymbol{0_4} \\  
~\boldsymbol{0_4} & ~+\frac{1}{2} \boldsymbol{\gamma}_{\hat{m}\hat{n}} 
\end{pmatrix}
\qquad (\hat{m},~\hat{n}=5,\ldots,9)\,, \no \\
\no \\ 
&& \left(\gQ^{\check{\alpha}\hat{\alpha}}\right)^{I} = \begin{pmatrix}
~\boldsymbol{0_4} & ~m_I{}^{\check{\alpha}\hat{\alpha}}\\  
~-\bar{m}_I{}^{\check{\alpha}\hat{\alpha}} & ~\boldsymbol{0_4}
\end{pmatrix}\,. 
\end{eqnarray}
Here we have used the following notation, 
\[
\boldsymbol{\gamma}_{\check{m}\check{n}} \equiv \frac{1}{2}[\boldsymbol{\gamma}_{\check{m}}, 
\boldsymbol{\gamma}_{\check{n}}]\,, \qquad 
\boldsymbol{\gamma}_{\hat{m}\hat{n}} \equiv \frac{1}{2}[\boldsymbol{\gamma}_{\hat{m}}, 
\boldsymbol{\gamma}_{\hat{n}}]\,.
\]
and introduced the following quantities, 
\begin{eqnarray}
(m_I{}^{\check{\alpha}\hat{\alpha}})_i{}^j&\equiv&
\text{e}^{(1-(-1)^I)i\,\frac{\pi}{4}}K^{j\check{\alpha}}\delta^{\hat{\alpha}}_i\,,\no\\
(\bar{m}_I{}^{\check{\alpha}\hat{\alpha}})_i{}^j&\equiv&
\text{e}^{(1+(-1)^I)i\,\frac{\pi}{4}}K^{\hat{\alpha}j}\delta^{\check{\alpha}}_i\,,
\end{eqnarray}
with the matrix $K$ defined as 
\begin{eqnarray}
K\equiv i\, \gamma_2\, \gamma_0=\begin{pmatrix}
~0 & -1 & ~0 & ~0 \\  
~1 & ~0 & ~0 & ~0 \\
~0 & ~0 & ~0 & -1 \\
~0 & ~0 & ~1 & ~0
\end{pmatrix}\,.
\end{eqnarray}

It is helpful to summarize the action of $d$ and $\tilde{d}$ defined in (\ref{op-d}) and (\ref{op-tilde-d}), 
respectively: 
\begin{eqnarray}
d(\gP_m)=\tilde{d}(\gP_m)=2\,\gP_m\,,\quad d(\gJ_{mn})=\tilde{d}(\gJ_{mn})=0\,,\quad 
d(\gQ^I)=-\tilde{d}(\gQ^I)=\sigma_3^{IJ}\gQ^J\,.
\end{eqnarray}

 \subsection*{Supertrace formulae} 

When the Lagrangian is evaluated, the following supertrace formulae are useful: 
\begin{eqnarray}
{\rm Str}\left[\,\gP_{m}\gP_{n}\,\right] &=& \eta_{mn}\,, \no \\ 
{\rm Str}\left[\,\gJ_{\check{m}\check{n}}\gJ_{\check{p}\check{q}}\,\right] 
&=& - (\eta_{\check{m}\check{p}}\eta_{\check{n}\check{q}} 
- \eta_{\check{m}\check{q}}\eta_{\check{n}\check{p}})\,, \no \\ 
{\rm Str}\left[\,\gJ_{\hat{m}\hat{n}}\gJ_{\hat{p}\hat{q}}\,\right]
&=& + (\eta_{\hat{m}\hat{p}}\eta_{\hat{n}\hat{q}} 
- \eta_{\hat{m}\hat{q}}\eta_{\hat{n}\hat{p}})\,, \no \\ 
{\rm Str}\left[(\gQ^{\check{\alpha}\hat{\alpha}})^I (\gQ^{\check{\beta}\hat{\beta}})^J \right]
&=& -2\epsilon^{IJ} K^{\check{\alpha}\check{\beta}}K^{\hat{\alpha}\hat{\beta}}\,. 
\end{eqnarray}

\subsection*{A lift up to the $16\times 16$ matrix representation}

Here let us consider a lift up of the $8\times 8$ matrix representation 
to the $16 \times 16$ one that appears 
in the $\mathfrak{su}(2,2|4)$ superalgebra (\ref{algebra})\,.

\medskip 

The $16\times 16$ gamma matrices $\gamma_m$ in the superalgebra (\ref{algebra}) 
can be realized as follows:  
\begin{eqnarray}
\gamma_m &=& (\gamma_{\check{m}}, \gamma_{\hat{m}}) \qquad (m=0,\ldots,9)\,, 
\end{eqnarray}
where $\gamma_{\check{m}}$ and $\gamma_{\hat{m}}$ are constructed as a tensor product 
with ${\bf 1_4}$ like 
\begin{eqnarray}
\gamma_{\check{m}} \equiv \boldsymbol{\gamma}_{\check{m}} \otimes {\bf 1_4}\,, \qquad 
\gamma_{\hat{m}} \equiv {\bf 1_4} \otimes i\, \boldsymbol{\gamma}_{\hat{m}}\,.
\end{eqnarray}
More explicitly, the index structure can be displayed as 
\begin{eqnarray}
&& \left(\gamma_m\right)_{\check{\alpha}\hat{\alpha}}^{~~~\check{\beta}\hat{\beta}} 
= \left( 
\left(\gamma_{\check{m}}\right)_{\check{\alpha}\hat{\alpha}}^{~~~\check{\beta}\hat{\beta}}, 
\left(\gamma_{\hat{m}}\right)_{\check{\alpha}\hat{\alpha}}^{~~~\check{\beta}\hat{\beta}}
\right)\,, \no \\ 
&& \left(\gamma_{\check{m}}\right)_{\check{\alpha}\hat{\alpha}}^{~~~\check{\beta}\hat{\beta}} 
= \left(\boldsymbol{\gamma}_{\check{m}}\right)_{\check{\alpha}}^{~~\check{\beta}} 
\otimes \left({\bf 1_4}\right)_{\hat{\alpha}}^{~~\hat{\beta}}\,, 
\qquad 
\left(\gamma_{\hat{m}}\right)_{\check{\alpha}\hat{\alpha}}^{~~~\check{\beta}\hat{\beta}} 
= \left({\bf 1_4}\right)_{\check{\alpha}}^{~~\check{\beta}} 
\otimes  i\, \left(\boldsymbol{\gamma}_{\hat{m}}\right)_{\hat{\alpha}}^{~~\hat{\beta}}\,. 
\end{eqnarray} 
Then the gamma matrices $\gamma_m$ act on the spinor 
$\theta_I = \left(\theta_{\check{\alpha}\hat{\alpha}}\right)_I$ like 
\begin{eqnarray}
\gamma_m \theta_I = 
\left(\gamma_m \right)_{\check{\alpha}\hat{\alpha}}^{~~~\check{\beta}\hat{\beta}} 
\left(\theta_{\check{\beta}\hat{\beta}}\right)_{I}\,. 
\end{eqnarray}

\subsection*{A lift up to the $32\times 32$ matrix representation}

In Sect.\ 3, it is necessary to rewrite the deformed Lagrangian in terms of 
ten-dimensional $32\times 32$ gamma matrices $\Gamma$ 
in order to read off the component fields of type IIB supergravity.
Hence we introduce a concise rule to switch from the $16\times 16$ notation 
to the $32\times 32$ one. 

\medskip 

Let us first define the following rules: 
\begin{eqnarray}
\bar{\Theta}_I \Gamma_m \Theta_J &\equiv& \bar{\theta}_I \gamma_m \theta_J\,,\no\\
\bar{\Theta}_I \Gamma_m \Gamma_{np} \Theta_J &\equiv &\bar{\theta}_I 
\gamma_m \gamma_{np} \theta_J\,.
\end{eqnarray}
Here we have defined $ \gamma_{mn}$ as 
\begin{eqnarray}
&&\gamma_{\check{m}\check{n}} \equiv \boldsymbol{\gamma}_{\check{m}\check{n}}
\otimes {\bf 1_4}\,,\no\\
&&\gamma_{\check{m}\hat{n}} \equiv 
\boldsymbol{\gamma}_{\check{m}}\otimes i\, \boldsymbol{\gamma}_{\hat{n}}\,,\no\\
&&\gamma_{\hat{m}\hat{n}} \equiv {\bf 1_4}\otimes \boldsymbol{\gamma}_{\hat{m}\hat{n}}\,.
\end{eqnarray}
Then the other combinations of gamma matrices are automatically lifted up as follows:
\begin{eqnarray}
&&\bth_I\,\gamma_m\,\gamma_n\,\theta_J= \bar{\Theta}_I\,\Gamma_m\,
\Gamma_{01234}\,\Gamma_n\,\Theta_J\,,\no\\
&&\bth_I\,\gamma_{mn}\,\theta_J=\bar{\Theta}_I\,\Gamma_{01234}\,\Gamma_{mn}\,\Theta\,,\no\\
&&\bth_I\,\gamma_{mn}\,\gamma_p\,\theta_J=\bar{\Theta}_I\,
\Gamma_{01234}\,\Gamma_{mn}\,\Gamma_{01234}\,\Gamma_p\,\Theta_J\,,\no\\
&&\bth_I\,\gamma_{mn}\,\gamma_{pq}\,\theta_J=\bar{\Theta}_I\,\Gamma_{01234}\,\Gamma_{mn}\,\Gamma_{pq}\,\Theta_J\,. 
\label{lift}
\end{eqnarray}
Note here that the right-hand side of (\ref{lift}) contains the factor $\Gamma_{01234}$, 
which is evaluated as 
\[
\Gamma_{01234}=\frac{1}{5!}\Gamma_{[0}\cdots\Gamma_{4]}
=\sigma_1\otimes {\bf 1_4}\otimes {\bf 1_4}\,.
\] 
The insertion of this factor is necessary for an appropriate lift-up. 
For the detail of the lift-up, see, for example, \cite{HKS}.

\end{document}